\documentstyle[preprint,aps,epsf]{revtex}

\newcommand {\bea}{\begin{eqnarray}}
\newcommand {\eea}{\end{eqnarray}}
\newcommand {\be}{\begin{equation}}
\newcommand {\ee}{\end{equation}}

\newcommand {\Dslash}{D\!\!\!/}
\newcommand {\dslash}{\partial\!\!\!/}

\begin{document}

\preprint{SUNY-NTG-00-07}

\title{Gluino Condensation in an Interacting Instanton 
Ensemble}

\author{Thomas Sch\"afer}

\address{Department of Physics, SUNY Stony Brook,
Stony Brook, NY 11794\\ and\\
Riken-BNL Research Center, Brookhaven National 
Laboratory, Upton, NY 11973}

\maketitle

\begin{abstract}
  We perform a semi-classical study of chiral symmetry
breaking and of the spectrum of the Dirac operator in QCD 
with adjoint fermions. For this purpose we calculate matrix 
elements of the adjoint Dirac operator between instanton zero 
modes and study their symmetry properties. We present simulations
of the instanton ensemble for different numbers of Majorana
fermions in the adjoint representation. These simulations 
provide evidence that instantons lead to gluino condensation 
in supersymmetric gluodynamics.

\end{abstract}

\section{Introduction}

 In order to improve our understanding of non-perturbative 
phenomena in QCD it is useful to view QCD from a larger 
perspective, as a member of a family of QCD-like theories
with different matter contents. In this context we would
like to understand the phase structure of QCD-like theories
with $N_f$ fermions in the fundamental representation of
the gauge group and $N_{ad}$ fermions in the adjoint 
representation. Theories with adjoint fermions are special
because the action may display a symmetry that connects 
bosonic and fermionic degrees of freedom, supersymmetry. 
Supersymmetry imposes powerful restrictions on the 
structure of the low energy effective action. These
constraints have been used, for example, to determine
the phase structure of $N=1$ supersymmetric QCD with
$N_c$ colors and $N_f$ flavors of fundamental quarks
\cite{Sei_94,IS_96}. 

 In addition to that, supersymmetry provides the 
opportunity to isolate certain non-perturbative effects, 
in particular instantons. This idea has been used in order 
to calculate the gluino condensate in the simplest supersymmetric
gauge theory, SUSY gluodynamics. The strategy behind
the so called weak coupling instanton (WCI) calculation
\cite{ADS_83,ADS_84,ADS_85,NSVZ_85,SV_88} is to add to 
the theory a fundamental fermion together with
its scalar superpartners and consider the regime where 
the expectation value of the scalar field is large. In
this case there is a unique non-perturbative 
superpotential induced by instantons. Since the scalar
vev is large, instantons are semi-classical and the 
superpotential can be calculated reliably. The 
superpotential determines the gluino condensate 
in the theory with additional matter. Finally, the 
matter fields can be decoupled by sending their
mass to infinity. The result for the gluino condensate
in $SU(2)$ supersymmetric gluodynamics is 
\be
\label{ll_wci}
\langle \lambda\lambda\rangle = 6\Lambda^3,
\ee
where $\Lambda$ is the scale parameter defined in
\cite{Shi_99}
\be 
 \Lambda = M_{PV} \left( \frac{16\pi^2}{bg^2}\right)^{1/3}
 \exp\left(-\frac{8\pi^2}{bg^2}\right).
\ee
Here, $b=3N_c$ is the first coefficient of the beta 
function in supersymmetric gluodynamics and $M_{PV}$
is a Pauli-Vilars regulator. 

 There is an old puzzle concerning this result. The
puzzle is connected with the fact that there is an 
alternative method for calculating the gluino condensate, 
usually referred to as the strong coupling instanton (SCI) 
method \cite{NSVZ_83,RV_84,ARV_85,FS_86}, for a review see 
\cite{AKM*88}. In the case of $N_c=2$ one considers a four
fermion correlation function. This correlation function is 
a topological quantity. Not only can it be saturated with
one instanton, but supersymmetry implies that the correlator 
is just a constant. At short distance, one expects that this 
constant is saturated by small instantons and can be calculated 
reliably. The gluino condensate is then extracted by 
using clustering. The puzzle is that the result differs 
from the WCI calculation by a factor 4/5.

 Several suggestions have been put forward in order 
to resolve the puzzle \cite{KS_97,DHKM_99,HKLM_99,VR_99}.
We do not wish to discuss these possibilities in detail. 
Instead, we would like to employ a somewhat different, 
more qualitative approach. Even though there is no
direct instanton contribution to the gluino condensate,
one would still expect configurations with instantons
and anti-instantons to contribute to the gluino 
condensate. Here we have in mind that the theory is 
studied in a finite volume and in the presence of
a non-zero mass term. The thermodynamic limit is approached 
by taking the volume to infinity before we let the mass
go to zero. The mechanism for gluino condensation is 
similar to the instanton liquid picture of quark condensation 
in ordinary QCD \cite{Shu_82,DP_85,DP_86}. For $N_f>1$ 
there is no direct instanton contribution to the quark 
condensate but chiral symmetry breaking may take place 
in an ensemble of instantons and anti-instantons in the 
thermodynamic limit. The Banks-Casher relation 
$\langle\bar qq\rangle=-\pi\rho(0)$ \cite{BC_80}
connects the quark condensate with the density of 
eigenvalues of the Dirac operator at zero virtuality.
For simplicity let us consider an ensemble with an 
equal number of instantons and anti-instantons. In 
this case the Dirac operator no longer has any exact
zero modes. However, if the interaction between the 
instantons is sufficiently weak, the approximate 
zero modes associated with individual instantons and
anti-instantons form a zone around zero virtuality 
and lead to spontaneous chiral symmetry breaking. 
This quark condensation mechanism has been investigated 
in some detail, both analytically and on the lattice 
\cite{SS_98,Neg_99}, and the results seem to support 
the instanton picture. 

  In the present work we wish to extend these studies
to theories with fermions in the adjoint representation.
Since we are dealing with a strongly coupled theory,
our calculations are necessarily approximate. In particular,
we will have to restrict ourselves to the contribution
of small instantons for which the semi-classical description 
is appropriate. On the other hand, the methods we are using  
are applicable also to non-supersymmetric theories with 
several flavors of adjoint fermions. In addition to that,
we can use these methods to study non-constant correlation
functions that determine the spectrum of the theory. 

 The paper is organized as follows. In section \ref{sec_adqcd}
we discuss some general aspects of chiral symmetry breaking 
in theories with fermions in the adjoint representation. In 
section \ref{sec_zm} we describe the structure of the instanton
zero mode wave functions and in \ref{sec_tia} we calculate
matrix elements of the Dirac operator between zero mode
states. These results are used in order to determine the 
fermion determinant in the field of an instanton-anti-instanton
pair (section \ref{sec_detia}) and to calculate the gluino
condensate in a random instanton ensemble (section \ref{sec_ran}).
In section \ref{sec_int} we describe simulations of an 
interacting instanton ensemble with different numbers 
of fermions in the fundamental and adjoint representation.

\section{QCD with adjoint fermions}
\label{sec_adqcd}

 QCD with adjoint fermions is defined by the lagrangian
\be
\label{l_adjqcd}
{\cal L} = \sum_{i=1}^{N_{ad}}\frac{1}{2}\bar\lambda_M^{(i)a}(i\Dslash)^{ab}
\lambda_M^{(i)b} -\frac{1}{4g^2} G_{\mu\nu}^aG_{\mu\nu}^a,
\ee
where $\lambda^a_M$ is a Majorana fermion in the adjoint representation
of the gauge group and $G^a_{\mu\nu}$ is the usual field strength 
tensor. The covariant derivative in the adjoint representation 
is given by
\be
\label{D_adj}
D_\mu^{ab} =\partial_\mu\delta^{ab} + f^{abc}A_\mu^c .
\ee
For several Majorana flavors the theory (\ref{l_adjqcd}) possesses
a $SU(N_{ad})$ chiral symmetry. A non-zero gluino condensate
\be
 \langle \bar\lambda_M^{(i)}\lambda_M^{(j)}\rangle
 = -\delta^{ij}\sigma
\ee
breaks this symmetry to $SO(N_{ad})$ \cite{KSV_85}. This fact can 
be seen most easily by considering the conserved vector and axial-vector 
currents \cite{LS_92}. There are $\frac{1}{2}N_{ad}(N_{ad}-1)$ 
conserved vector currents $V^{ij}_\mu=\bar\lambda_M^{(i)}\gamma_\mu
\lambda_M^{(j)}$ and $\frac{1}{2}N_{ad}(N_{ad}+1)$ classically
conserved axial-vector currents $A^{ij}_\mu=\bar\lambda_M^{(i)}
\gamma_\mu\gamma_5\lambda_M^{(j)}$. The singlet axial current 
$A^{ii}_\mu$ is anomalous. At the quantum level this leaves
$N_{ad}^2-1$ conserved charges that generate the $SU(N_{ad})$
chiral symmetry. Gluino condensation breaks the axial symmetries
and leads to the appearance of $\frac{1}{2}(N_{ad}^2+N_{ad}-2)$
Goldstone bosons. The unbroken $\frac{1}{2}N_{ad}(N_{ad}-1)$ vector 
charges generate the residual $O(N_{ad})$ symmetry. 

 In the case of supersymmetric gluodynamics, $N_{ad}=1$, there
is no continuous symmetry. Instantons break the axial $U(1)_A$
symmetry but leave a discrete $Z_{N_c}$ symmetry intact. This
discrete symmetry is spontaneously broken by gluino condensation. 
As discussed above, the value of the gluino condensate is 
known from a weak coupling instanton calculation. There are
no predictions for the spectrum of the theory, but we expect
the lowest states to fill out a chiral supermultiplet containing
a scalar and a pseudoscalar meson as well as a Majorana fermion.
These results can be summarized in terms of an effective 
lagrangian \cite{VY_82}. This is not an effective lagrangian
in the Wilsonean sense. The effective action does not generate
the low momentum scattering amplitudes of the theory. Instead,
it mainly serves as a generating functional for the anomalous
Ward identities of the theory.

\section{Instanton gauge potential and fermionic zero modes}
\label{sec_zm}

  In theories with adjoint fermions it is convenient to 
employ a spinor notation for spin, vector, and color
indices \cite{Shi_99}. We can convert vectors to spinors using
\be
V_{\alpha\dot\alpha} = V_\mu (\sigma_\mu)_{\alpha\dot\alpha}.
\ee
The euclidean spinor conventions used in this paper are
summarized in Appendix A. The instanton gauge potential couples 
spin to color degrees of freedom. A field $A^a$ in the adjoint 
representation of $SU(2)$ can be represented by a symmetric 
tensor $A^{\alpha
\beta}$
\be
A^a = A^{\alpha\beta}\epsilon_{\alpha\gamma}(\tau^a)^\gamma_\beta .
\ee
In spinor notation, the instanton gauge potential 
in regular gauge is given by
\be
 A^{\alpha\beta}_{\gamma\dot\delta} 
 = -2i(\delta^\alpha_\gamma x^\beta_{\dot\delta}
   +\delta^\beta_\gamma x^\alpha_{\dot\delta})
   \frac{1}{x^2+\rho^2} .
\ee
We can transform the gauge potential to singular gauge using
the gauge transformation 
\be
U^{\dot\alpha\alpha}=\hat{x}_\mu(\bar\sigma_\mu)^{\dot\alpha\alpha}.
\ee
Note that this matrix transforms an undotted color index into
a dotted one. We can perform a `fake' conversion of the dotted
spinor back to an undotted one using the fact that $(\sigma_0)^{
\alpha\dot\alpha}$ is just the unit matrix. 

 In the case of $SU(2)$, the Dirac operator in the background
field of an instanton has four zero modes. The first two are
conventionally called the supersymmetric (ss) zero modes
\cite{AKM*88}
\be
\lambda^{\gamma\delta}_{\alpha(\beta)} = 
 (\delta^\gamma_\alpha\delta^\delta_\beta
 +\delta^\gamma_\beta\delta^\delta_\alpha)
 \frac{\rho^2}{\pi}\frac{1}{(x^2+\rho^2)^2} ,
\ee
where $\beta=1,2$ enumerates the zero modes. The other two
are referred to as the superconformal (sc) zero modes
\be
\lambda^{\gamma\delta}_{\alpha(\dot\beta)} = 
 (\delta^\gamma_\alpha x^\delta_{\dot\beta}
 +\delta^\delta_\alpha x^\gamma_{\dot\beta})
 \frac{\rho}{\sqrt{2}\pi}\frac{1}{(x^2+\rho^2)^2}.
\ee
In singular gauge, the zero modes are given by
\bea
\label{ss_sing}
\lambda^{\dot\gamma\dot\delta}_{\alpha(\beta)} = 
 (x^{\dot\gamma}_\alpha x^{\dot\delta}_\beta
 +x^{\dot\gamma}_\beta x^{\dot\delta}_\alpha)
 \frac{\rho^2}{\pi}\frac{1}{x^2(x^2+\rho^2)^2}
   \hspace{0.5cm}(ss) , \\
\label{sc_sing}
\lambda^{\dot\gamma\dot\delta}_{\alpha(\dot\beta)} = 
 (x^{\dot\gamma}_\alpha \delta^{\dot\delta}_{\dot\beta}
 +x^{\dot\delta}_\alpha \delta^{\dot\gamma}_{\dot\beta})
 \frac{\rho}{\sqrt{2}\pi}\frac{1}{(x^2+\rho^2)^2}
\hspace{0.5cm}(sc).
\eea
Analogously, we can construct the zero modes of the 
Dirac operator in the background field of an anti-instanton. 
The regular gauge supersymmetric zero mode has the 
structure 
$\lambda^{\dot\alpha(\dot\beta)}_{\dot\gamma\dot\delta}
\sim(\delta^{\dot\alpha}_{\dot\gamma}\delta^{\dot\beta}_{\dot\delta} 
    +\delta^{\dot\alpha}_{\dot\delta}\delta^{\dot\beta}_{\dot\gamma})$,
etc. 

 The effect of the zero modes on the propagation of fermions 
can be summarized in terms of an effective lagrangian \cite{tHo_76}.
The 't Hooft effective interaction in the case of one Majorana 
fermion in the adjoint representation of $SU(2)$ was determined 
in \cite{VZ_82,NSVVZ_83}. The result is 
\bea 
\label{l_tHo}
{\cal L} &=& \frac{4\pi^4}{3}\left(\frac{2\pi}{\alpha_s}\right)^4
 \exp\left(-\frac{2\pi}{\alpha_s}\right) \rho^3 d\rho \,
\left\{ \bar\lambda_M^a\lambda_M^a \partial_\mu\bar\lambda_M^b
 \partial^\mu\lambda_M^b
 + \bar\lambda_M^a\gamma_5\lambda_M^a \partial_\mu\bar\lambda_M^b
 \gamma_5\partial^\mu\lambda_M^b \right.
\nonumber \\
& &\hspace{4cm}\mbox{}\left.
 - \frac{1}{2} \bar\lambda_M^a\sigma_{\alpha\beta}\lambda_M^b
   \partial_\mu\bar\lambda_M^b\sigma^{\alpha\beta}\partial^\mu\lambda_M^a
 \right\}.
\eea
This result has to be interpreted with some care. The notion of 
an effective interaction induced by instantons of some fixed size 
is incompatible with supersymmetry. In order to derive 
manifestly supersymmetric results we always have to integrate
over the collective coordinates of the instanton. Nevertheless,
it is instructive to compare the result (\ref{l_tHo}) with 
the effective interaction in the case of $N_f=2$ Dirac fermions 
in the fundamental representation. The structure of the two
interactions is quite similar, which suggests that instantons
may lead to similar physical effects. The most important 
difference between the two effective lagrangians is the presence
of derivatives acting on two of the four Majorana spinors in
(\ref{l_tHo}). This difference is connected with the asymptotic
behavior of the supersymmetric zero modes, which is not $\sim
1/z^3$, but $1/z^4$.

\section{Matrix elements of the Dirac operator}
\label{sec_tia}

 In the following, we wish to study the spectrum of the Dirac
operator in an instanton ensemble. For this purpose, we have
to calculate matrix elements of the Dirac operator between
the zero modes of individual instantons and anti-instantons
\be
\label{tia}
 T_{IA} = \int d^4x \, \bar\lambda^a_I (i\Dslash)^{ab}\lambda^b_A.
\ee
An ensemble of instantons and anti-instantons is only an 
approximate saddle point of the action. If the system is
sufficiently dilute then the instantons and anti-instantons
are well separated and the approximate saddle point solution
for the gauge potential is given by a simple sum of the 
gauge potentials of the individual instantons. For this 
purpose, the gauge potential of the individual instantons
has to be put in singular gauge. In the sum ansatz, we 
can use the equations of motion of the fermion fields in
order to replace the covariant derivative in (\ref{tia})
by an ordinary derivative
\be
\label{tia2}
 T_{IA} = -\int d^4x \, \bar\lambda^a_I (i\dslash)\lambda^a_A.
\ee
The structure of the Dirac operator is dictated by the form of
the zero modes. In the background field of an instanton-anti-instanton
pair we have
\be
\label{tia_sym}
T_{IA} = \left(
\begin{array}{cc}
 0 &
\begin{array}{cc}
T^{ss-ss}_{IA} & T^{ss-sc}_{IA} \\
T^{sc-ss}_{IA} & T^{sc-sc}_{IA}
\end{array} \\
\begin{array}{cc}
T^{ss-ss}_{AI} & T^{ss-sc}_{AI} \\
T^{sc-ss}_{AI} & T^{sc-sc}_{AI}
\end{array} & 0
\end{array}\right),
\ee
where the matrix elements $T^{ss}_{AI},\ldots$ are real 
quaternions. These quaternions can be decomposed as
\bea
\label{tia_comp}
(T^{ss-ss}_{AI})_{\dot\beta\beta'} &=&
  T^{ss}_\mu (\sigma_\mu)_{\beta'\dot\beta} \\
(T^{sc-sc}_{AI})_{\beta\dot\beta'} &=&
  T^{sc}_\mu (\sigma_\mu)_{\beta\dot\beta'} \\
(T^{ss-sc}_{AI})_{\dot\beta\dot\beta'}  &=&
  T^{ss-sc}\epsilon_{\dot\beta\dot\beta'} +
  T^{ss-sc}_{\mu\nu} \epsilon_{\dot\beta\dot\gamma}
(\bar\sigma_{\mu\nu})^{\dot\gamma}_{\;\dot\beta'},\\
(T^{sc-ss}_{AI})_{\beta\beta'}    &=&
  T^{sc-ss}\epsilon_{\beta\beta'} +
  T^{sc-ss}_{\mu\nu} (\sigma_{\mu\nu})_\beta^{\;\gamma}
  \epsilon_{\gamma\beta'} .
\eea
Here, $T^{ss}_\mu$ and $T^{sc}_\mu$ are real vectors,
$T^{ss-sc}$ and $T^{sc-ss}$ are real scalars and 
$T^{ss-sc}_{\mu\nu}$ and $T^{sc-ss}_{\mu\nu}$ are
self-dual and anti-self-dual tensors, respectively. 
Chiral symmetry implies that the diagonal blocks 
of $T_{IA}$ are zero. The upper right and lower 
left blocks are related by hermitean conjugation. 
For example, we find that 
\be
(T_{IA}^{ss-ss})_{\beta\dot\beta'} = T^{ss}_\mu
 (\bar\sigma^\mu)^{\dot\beta'\beta}.
\ee
In general, we have $(T^\dagger)_{AI} = (T)_{IA}$. 
The eigenvalues of (\ref{tia_sym}) come in quartets
$(\xi,\xi,-\xi,-\xi)$. These results are in agreement 
with the general arguments presented in \cite{LS_92,Ver_94}. 

 The functions $T^{ss}_\mu,\ldots$ depend on the collective
coordinates of the instanton and anti-instanton. We will 
characterize the relative color orientation by the four 
vector $u_\mu=1/2\cdot{\rm tr}(U_A^\dagger U_I\sigma_\mu)$.
Here, $U_{I,A}$ are $SU(N_c)$ matrices that characterize 
the color orientation of the instanton and anti-instanton. 
For color $SU(2)$ $u_\mu$ is a real vector with $u^2=1$. 
Using rotational symmetry and the fact that $T_{AI}$ is
quadratic in $u_\mu$ we have
\bea
T^{ss}_\mu\;\;\; &=& \hat{z}_\mu T^{ss}_1 + u_\mu (u\cdot\hat{z})
  T^{ss}_2 + \hat{z}_\mu (u\cdot\hat{z})^2 T^{ss}_3,\\
T^{sc}_\mu \;\;\; &=& \hat{z}_\mu T^{sc}_1 + u_\mu (u\cdot\hat{z})
  T^{sc}_2 + \hat{z}_\mu (u\cdot\hat{z})^2 T^{sc}_3,\\
T^{ss-sc} & =&  T^{ss-sc}_1 + T^{ss-sc}_2 (u\cdot\hat{z})^2,\\
T^{ss-sc}_{\mu\nu} & =& (u_\mu\hat{z}_\nu - u_\nu\hat{z}_\mu)
                (u\cdot\hat{z}) T^{ss-sc}_3,
\eea
where $z_\mu=z^A_\mu-z^I_\mu$ and the functions $T^{ss}_1,\dots$
depend on $(|z_\mu|,\rho_I,\rho_A)$. For simplicity, we will 
assume that the dependence on $\rho_{I,A}$ only enters through
their geometric mean $\bar\rho=\sqrt{\rho_I\rho_A}$. The fact 
that this assumption is valid to fairly good accuracy was 
checked in the case of fundamental fermions. In a more 
sophisticated treatment of the instanton-anti-instanton
gauge configuration the dependence of the overlap matrix 
element on $(z,\rho_I,\rho_A)$ is restricted by conformal
invariance \cite{Ver_92}. 

 In the following we shall outline the calculation of the 
invariant functions $T^{ss}_1,\ldots$. We describe the case
$T^{ss}_1$ in some detail but relegate the results for the
other functions to appendix B. Using the expression (\ref{ss_sing})  
for the wave function of the supersymmetric zero modes in 
singular gauge we find
\bea
\label{tss_full}
T^{ss}_\eta &=& \Big\{ 
{\rm tr}\left(\bar\sigma_\mu \sigma_\rho 
       \bar\sigma_\beta \sigma_\alpha\right)
{\rm tr}\left(\bar\sigma_\nu \sigma_\sigma 
       \bar\sigma_\gamma\sigma_\eta\right)
 + {\rm tr}\left(\bar\sigma_\mu \sigma_\rho \bar\sigma_\beta
 \sigma_\eta \bar\sigma_\nu \sigma_\sigma \bar\sigma_\gamma
 \sigma_\alpha \right) \Big\}
\nonumber \\
& & \hspace{5cm}\cdot
 u_\rho u_\sigma 
 \int d^4x \,\phi_{\mu\nu}(x-z)\partial_\alpha 
 \phi_{\beta\gamma}(x),
\eea
where $\phi_{\mu\nu}(x)$ is the profile function of the 
supersymmetric zero mode
\be
\label{phimunu} 
\phi_{\mu\nu}(x) = \frac{\rho^2}{\pi} 
 \frac{\hat{x}_\mu\hat{x}_\nu}{(x^2+\rho^2)^2}.
\ee
The integral in (\ref{tss_full}) is most easily calculated
in Fourier space. The Fourier transform of $\phi_{\mu\nu}$
is given by
\be
\label{phiss_ft}
\phi_{\mu\nu}(k) = \delta_{\mu\nu}\phi_1(k) 
  + \hat{k}_\mu\hat{k}_\nu \phi_2(k) 
\ee
with
\bea
\phi_1(k) &=& \;\, \frac{2\pi\rho^2}{y} \left\{
 \frac{4}{y^3} - \left(\frac{4}{y^2}+1\right)K_1(y)
 -\frac{2}{y} K_0(y)\right\}, \\
\phi_2(k) &=& - 2\pi\rho^2 \left\{
 \frac{16}{y^4} - \left(\frac{16}{y^3}+\frac{4}{y}\right)K_1(y)
 -\left(\frac{8}{y^2}+1 \right) K_0(y)\right\},
\eea
and $y=k\rho$. $K_n(y)$ is the modified Bessel function of 
the first kind and order $n$. We can now calculate the overlap 
integral and perform the traces. In momentum space the result 
is given by
\be 
T^{ss}_\eta (k) = (-i)\left( -2\hat{k}_\eta - 8u_\eta (u\cdot\hat{k})
 +16 (u\cdot\hat{k})^2 \right) |\phi_2(k)|^2.
\ee
Finally, we can determine the functions $T^{ss}_{1,2,3}$ by
performing the inverse Fourier transform. In the $d^4k$ integral
all integrations except for the one over the absolute magnitude 
of $k$ can be performed analytically. We find
\bea
\label{tss_z}
T^{ss}_1(z) &=& \frac{1}{8\pi^2} \int dk\,
 \left\{ 2k^4j_1(kz) -16 \frac{k^3}{z} j_2(kz)\right\}
 |\phi_2(k)|^2 \\
T^{ss}_2(z) &=& \frac{1}{8\pi^2} \int dk\,
 \left\{ 8k^4j_1(kz) -32 \frac{k^3}{z} j_2(kz)\right\}
 |\phi_2(k)|^2 \\
T^{ss}_3(z) &=& \frac{1}{8\pi^2} \int dk\,
 \; 16 k^4 j_3(kz) \, |\phi_2(k)|^2 ,
\eea
where $j_n(x)$ is the spherical Bessel function of order $n$.
The integrals (\ref{tss_z}) have to be performed numerically.
The results are shown in Figure 1. In the following, we will
use a simple parameterization of the numerical results. In the 
case of the supersymmetric overlaps, we use
\bea
\bar{\rho}T^{ss}_1(z) &=&  \frac{-1.26\bar{z}}
 {1.0+2.34\bar{z}^2+0.35\bar{z}^4+0.24\bar{z}^6}  ,\\
\bar{\rho}T^{ss}_2(z) &=& \frac{1.05\bar{z}}{(1.0+0.38\bar{z}^2)^3} 
 + \frac{-6.36\bar{z}^3}{(1.0+0.68\bar{z}^2)^4}  ,\\
\bar{\rho}T^{ss}_3(z) &=& \frac{15.8\bar{z}^3}
 {(1.0+0.84\bar{z}^2)^4},
\eea
where $\bar{z}=z/\bar{\rho}$.
These parameterizations respect the asymptotic behavior of
the overlap integrals. In particular, we have $T^{ss}(z)\sim
1/z^5$, $T^{sc}(z)\sim 1/z^3$ and $T^{ss-sc}(z)\sim 1/z^4$.

  For completeness, let us compare these results to the 
corresponding expressions in the case of fundamental fermions.
In this case, there is only one zero mode per instanton. 
The overlap matrix element $T_{IA}$ is a real number in 
the case of $SU(2)$, and complex for $SU(N_c>2)$. $T_{IA}$
satisfies the symmetry relation $T_{IA}=T^*_{AI}$. As a
consequence, the eigenvalues are real and occur in pairs
$(\xi,-\xi)$. We can extract the dependence of $T_{IA}$
on the collective coordinates. The result is 
\be
\label{T_fund}
T^{fund}_{AI} = (u\cdot\hat{z})T^f(z,\rho_I,\rho_A),
\hspace{1cm}
T^f(z,\rho_I,\rho_A)\simeq \frac{1}{\rho_I\rho_A}
\frac{4z}{(2.0+z^2/(\rho_I\rho_A))^2}.
\ee
We note that the fundamental overlap matrix element
only depends on one $SU(2)$ angle $\cos\theta \equiv
(u\cdot\hat{z})$. From the asymptotic form of the zero
mode solution one finds $T^f(z)\sim 1/z^3$.
 
\section{The fermion determinant in the field of an 
instanton-anti-instanton pair}
\label{sec_detia}

 Before we study gluino condensation we would like to 
make a brief digression and discuss the gluino induced 
instanton-anti-instanton interaction. This interaction 
will play an important role in the calculation of the 
gluino condensate in an interacting instanton ensemble.
 
  The probability to find an instanton-anti-instanton pair
characterized by the collective coordinates $(z_{I,A},
\rho_{I,A},U_{I,A})$ is controlled by the weight factor
$\exp(-S) \det(\Dslash+m)$. The first factor is the 
well known gluonic interaction. If the instanton and
anti-instanton are well separated it has the dipole
form \cite{CDG_78}
\be
\label{s_int}
 S = 2S_0 - S_0\frac{4\rho_I^2\rho_A^2}{z^4}
 \left( 1-4\cos^2\theta \right),
\ee
where $S_0=(8\pi^2)/g^2$ is the single instanton action
and $\cos\theta$ is the $SU(2)$ angle introduced above. 
We note that the interaction is attractive if the color
orientation is aligned with the spatial orientation, 
$\cos\theta=\pm 1$. The second factor is the fermion 
determinant. In the case of fundamental fermions, it is 
also well known. We have 
\be
\label{det_f}
\det(\Dslash ) = \cos^2\theta \frac{16}{\rho_I^2\rho_A^2}
 \frac{z^2}{\left(2.0+z^2/(\rho_I\rho_A)\right)^4},
\ee
which is also attractive for $\cos\theta=\pm 1$. We also
note that the interaction peaks at $z^2\simeq\rho_I\rho_A$. 

  Using the results of the last section we can calculate 
the fermion induced interaction with fermions in the adjoint 
representation. In order to calculate the determinant for
one Majorana fermion we take the square root of the 
corresponding expression for a Dirac fermion in the 
adjoint representation. For simplicity, let us begin with
the determinant in the basis of the supersymmetric zero
modes only. We find
\bea
\label{det_susy}
\det(\Dslash)^{ss} &=& \left|  (T_1^{ss})^2 
 + \left( (T_2^{ss})^2 +2T_1^{ss}T_2^{ss} +2T_1^{ss}T_3^{ss}\right)
  \cos^2\theta \right.\\
& &\hspace{4cm} \mbox{} \left.
 + \left( (T_3^{ss})^2 +2T_2^{ss}T_3^{ss} \right)
  \cos^4\theta \right| . \nonumber 
\eea
The result for the superconformal zero modes is even more simple,
\be
\label{det_suco}
\det(\Dslash)^{sc} = \left|  (T_1^{sc})^2 
 + \left( (T_2^{sc})^2 +2T_1^{sc}T_2^{sc} \right)  \cos^2\theta
 \right| .
\ee
This expression is quite similar to the determinant for fundamental
fermions. The supersymmetric determinant (\ref{det_susy}) is somewhat 
more complicated, but also peaked for $\cos\theta=\pm 1$. When the 
mixing between supersymmetric and superconformal zero modes is
included the fermion determinant depends on other $SU(2)$ angles 
in addition to $\cos\theta$. We show numerical results for 
$\log(\det(\Dslash))$ as a function of $z$, $\cos\theta$ and $\cos\phi$
in Fig.\ref{fig_det}. Here we have taken $\hat{z}_\mu=z\delta_{\mu 4}$
and defined $\cos\theta = u_4$ and $\sin\theta\cos\phi=u_2$.
We observe that again the determinant peaks for $z^2\simeq
\rho_I\rho_A$. For large $z$, the determinant behaves as
$\sim 1/z^{16}$. More importantly, we find that the 
interaction is again most attractive for $\cos\theta=\pm 1$. 
There is some dependence on $\cos\phi$, but it is not as 
pronounced as the dependence on $\cos\theta$. This means that 
the gluino induced interaction for one Majorana fermion is 
qualitatively similar to the quark induced interaction with 
an effective number of quark flavors between $N_f=2$ (which 
gives $\det\sim 1/z^{12}$) and $N_f=3$ (corresponding to 
$\det\sim 1/z^{18}$).

\section{Gluino condensation in a random instanton ensemble}
\label{sec_ran}

 In this section we study gluino condensation in a
random instanton ensemble. This means that we will 
assume that the collective coordinates of the instantons
and anti-instanton are distributed randomly. In 
particular, we shall neglect the effect of the fermion
determinant on the distribution of instantons. This is 
not a good approximation even in ordinary QCD and it 
certainly cannot be correct in a supersymmetric theory.
Nevertheless, using the approximation of randomness
we can get some analytic understanding of the dependence
of the gluino condensate on the parameters characterizing
the instanton liquid. We can also get an estimate of
the relative size of the quark and gluino condensates 
in theories with both fundamental and adjoint fermions. 

 The simplest model of the spectrum of the Dirac operator
is based on the assumption that the non-zero matrix elements
of the Dirac operator are Gaussian random numbers \cite{DP_85}. 
The distribution is characterized by the first moment
\be 
 \sigma^2 = \langle \frac{2}{N}{\rm tr}\left(T^\dagger T
\right)\rangle .
\ee
The eigenvalue distribution for the Gaussian ensemble is
given by a semi-circle where the density of eigenvalues at 
zero virtuality is $\rho(0)=(N/V)(\pi\sigma)^{-1}$. Here, 
$(N/V)$ is the number of eigenstates per unit volume. 
The first moment of the overlap matrix can be estimated
by averaging $|T_{AI}|^2$ over the collective coordinates
of the instantons. Using (\ref{T_fund}) we find the first
moment of the Dirac operator for fermions in the fundamental
representation of $SU(2)$ \cite{DP_85}
\be
\label{sig_fund}
 \sigma = \left(\frac{1}{3}\frac{N}{V}\right)^{1/2}
 \bar\rho\pi ,
\ee
where $\bar\rho$ is the average size of the instanton. 
This parameter, just like the density of instantons,
cannot be determined in the semi-classical approximation.
In the instanton liquid model of the QCD vacuum it is 
assumed that $\bar\rho=1/3$ fm and $(N/V)=1\,{\rm fm}^4$
\cite{Shu_82}. Using these values we find
\be 
\label{qq_f_ran}
\langle\bar{q}q\rangle = -\frac{1}{\pi\bar\rho}
 3^{1/2}\left(\frac{N}{V}\right)^{1/2}
 \simeq -(230\,{\rm MeV})^3,
\ee
in very good agreement with the phenomenological value
(which, of course, applies to color $SU(3)$).

 The same arguments can be applied to gluino condensation
in a random instanton ensemble. In this case we need to 
determine the first moment of a quaternionic matrix with
the matrix elements determined in section \ref{sec_tia}. 
We find
\be
\label{sig_ad}
 \sigma_{ad} = \left(\frac{N}{V}\right)^{1/2}
  0.43 \bar\rho\pi ,
\ee
which is somewhat smaller than (\ref{sig_fund}). There
are four times as many eigenstates per unit volume
but for a Majorana fermion the Banks-Casher relation
has an additional factor 1/2, $\langle\bar\lambda
\lambda\rangle = - \pi/2\cdot\rho(0)$. We finally get the 
following estimate of the gluino condensate 
\be
\label{qq_ad_ran}
\langle \bar\lambda\lambda\rangle =-\frac{1}{\pi\bar\rho}
4.6 \left(\frac{N}{V}\right)^{1/2}. 
\ee
Here and in what follows we have dropped the subscript
M on the Majorana spinors. This result is a little more 
than twice as large as the corresponding result for a Dirac 
fermion in the fundamental representation. As we saw, this 
is mainly due to the effective number of zero modes in both 
cases. We emphasize that the gluino condensate is 
proportional to the square root of the instanton density,
which is also what one would expect if the condensate 
is extracted from the four-point function using clustering.

 We have checked these estimates by performing a 
numerical calculation of the spectrum of the Dirac 
operator in a random instanton ensemble. This means 
that instead of assuming the matrix elements of the 
Dirac operator to be random we take the collective 
coordinates of the instantons and anti-instantons to 
be random. We calculate the spectrum of the Dirac
operator and determine the gluino condensate using
\be
\langle\bar\lambda\lambda\rangle =
-\frac{1}{2}\int d\lambda\,\rho(\lambda)\,
 \frac{2m}{\lambda^2+m^2}.
\ee
The results are shown in Fig. \ref{fig_spec}. We observe 
that the spectrum is not a semi-circle but is peaked 
towards zero virtuality \cite{OV_98}. This 
non-analyticity is smoothed out when we calculate
the condensate for a non-zero quark or gluino 
mass. Again using $(N/V)=1\,{\rm fm}^4$ and 
$\bar\rho=1/3\,{\rm fm}$ as well as $m_q=m_g
=20$ MeV we find $\langle\bar\psi\psi\rangle = 
-(260\,{\rm MeV})^3$ and $\langle\bar\lambda
\lambda \rangle= -(347\,{\rm MeV})^3$.  

\section{Gluino condensation in an unquenched instanton
ensemble}
\label{sec_int}

 As we stressed in the previous section, the assumption
of randomness is not expected to be very useful. The 
fermion determinant is given by the product of all
eigenvalues of the Dirac operator, while the quark or
gluino condensate is determined by the density of 
small eigenvalues. This implies that the determinant
tends to suppress fermion condensates. In particular,
we expect that the strength of chiral symmetry 
breaking is reduced as the number of fermion flavors
is increased. 

 In this section we shall study this problem using 
simulations of the instanton ensemble in QCD with
fundamental and adjoint fermions. We consider the 
partition function
\be
\label{part}
Z = \int \left(\prod_i^N d\Omega_i d(\rho_i)\right)\, 
 \det(\Dslash_f+m_q)^{N_f}\det(\Dslash_a+m_g)^{N_{ad}/2} \exp(-S)
\ee
Here, $\Omega$ denotes the collective coordinates of 
the instanton, $d(\rho)$ is the single instanton
distribution \cite{tHo_76,NSVVZ_83,NSVZ_83b}, $\Dslash_{f,a}$
are the Dirac operators in the fundamental and adjoint
representation, and $\exp(-S)$ is the gluonic interaction
between instantons. In order to study spontaneous symmetry
breaking in a finite volume we introduce non-zero quark
and gluino masses $m_{q,g}$. We will study the limit 
$m_{q,g}\to 0$ in some detail. 

 The partition function (\ref{part}) suffers from the 
usual IR problem connected with large instantons for
which the semi-classical approximation does not apply. 
In practice, we deal with this problem by introducing
a short range repulsive core in the gluonic instanton
interaction, see section V.C. in \cite{SS_98} for a 
more detailed discussion. The repulsive core eliminates
the contributions of large instantons and very close 
pairs. This particular method for suppressing objects
that are not semi-classical has the virtue that it
respects the classical scale invariance of Yang-Mills
theory. 

 The instanton ensemble is characterized by two numbers,
the scale parameter $\Lambda$ that enters into the 
instanton weight $d(\rho)$ and a dimensionless parameter
$A$ which determines the size of the core. Lacking a 
better theory of topological fluctuations beyond the 
semi-classical domain we have to fix $A$ phenomenologically. 
This could be done, for example, as soon as lattice 
information on the spectrum and other properties of
theories with adjoint fermions becomes available
\cite{Cam_99,Kir_99}. In this work we will use the 
same value that was employed in studies of QCD with
fundamental fermions. It leads to a dilute instanton
ensemble characterized by the dimensionless 
parameter $\bar\rho^4(N/V)\simeq 0.12$. For
simplicity we will concentrate on simulations 
at a fixed instanton density $(N/V)=1.0\,\Lambda^4$.

 To set the stage, we show results for $N_f=1,\ldots,4$
flavors of fundamental fermions. Fig. \ref{fig_qq} shows
the quark condensate as a function of the quark mass
from simulations in a euclidean box of size $V=2.0^4 \Lambda^4$.
The case of only one flavor is special. The chiral condensate
persists even if the limit $m_q\to 0$ is taken in a finite
volume.
This is due to the fact
that for $N_f=1$ the quark condensate is dominated by direct
instanton contributions. The result for $N_f=2$ is characteristic
of spontaneous symmetry breaking. The quark condensate 
vanishes as the quark mass goes to zero but shows a clear
plateau for larger quark masses. One can verify that the 
onset of chiral symmetry breaking moves towards smaller 
masses as the volume is increased. For more than two flavors
the chiral condensate is significantly reduced. In the case
of three flavors the signal is already quite weak. Using
simulations in bigger volumes one can verify that chiral
symmetry is indeed broken. There is no clear evidence 
for chiral symmetry breaking in simulations with four or
more flavors. 

 Fig. \ref{fig_qq}b shows the gluino condensate measured
in simulations with one or two flavors of Majorana fermions
in the adjoint representation. For $N_{ad}=1$ there is 
clear evidence for spontaneous symmetry breaking. Indeed,
the behavior is more reminiscent of the case $N_f=1$, where
$\langle\bar qq\rangle$ receives direct instanton contributions,
than the case $N_f=2$, in which chiral symmetry breaking is a
collective effect. 


 These observations can be understood in more physical terms.
Supersymmetric gluodynamics has no Goldstone bosons, so finite
volume effects are much weaker than in $N_f=2$ non-supersymmetric
QCD. This means that in a fixed volume, gluino condensation can
be observed for gluino masses that are significantly smaller
than the quark masses required to produce quark condensation. 
In the standard picture, 
there is a discrete chiral symmetry which is broken by gluino
condensation. This means that if the gluino mass is too small
then chiral symmetry will be restored because of tunneling
between the $Z_2$ vacua. This is different from $N_f=1$ 
non-supersymmetric QCD where instantons leave no unbroken
discrete symmetries. 

 The value of the gluino condensate is $\langle\bar\lambda
\lambda\rangle \simeq 2\Lambda^3$. This result has the correct
order of magnitude but it cannot yet be compared directly to 
the prediction (\ref{ll_wci}). First of all, we use a 
different definition of the scale parameter. In order to
make contact with our work on QCD we use a Pauli-Vilars
scale parameter. Second, we have an additional parameter
$A$ which controls the boundary of the semi-classical
regime. Finally, we have performed the simulations at 
a fixed density of instantons $(N/V)=1.0\,\Lambda^4$. It is 
this choice which effectively sets the scale in our 
calculation.

 In Fig. \ref{fig_qq}b we also show the gluino condensate 
measured in simulations with $N_{ad}=2$ Majorana flavors. 
The condensate is very small and there is no clear evidence
for spontaneous chiral symmetry breaking. 

 The spectrum of the Dirac operator for $N_f=2$ quark flavors 
and $N_{ad}=1$ Majorana flavor is shown in Fig. \ref{fig_speci}a 
and b. The spectra were determined in simulations with $m_{q,g}
=0.1\Lambda^{-1}$. Again, we observe that in both cases there 
is a finite density of eigenvalues as $\lambda\to 0$. For 
$N_f=2$ the spectral density near $\lambda=0$ is flat\footnote{
Fig. \ref{fig_speci} shows that the spectral density is flat
for intermediate values of $\lambda$. There is a finite volume 
suppression of the spectral density for small $\lambda$ and a
$O(m^2)$ peak at $\lambda=0$. To show that the spectral density 
is flat at $\lambda=0$ in the limit $V\to \infty,m\to 0$
requires more numerical work.}, whereas
in the case $N_{ad}=1$ it is growing towards small $\lambda$.
Again, this is similar to the case of only one fundamental 
fermion. The results are consistent with the effective field
theory prediction \cite{SS_93,TV_99}
\be
\label{slope}
\rho'(\lambda\!=\!0)=\frac{\Sigma_0^2}{16\pi^2f_\pi^4}
  \frac{(N_f-2)(N_f+\beta)}{\beta N_f}.
\ee
Here, $\beta$ is the Dyson index of the random matrix ensemble
with the appropriate symmetry. We have $\beta=1$ for fundamental
fermions in $SU(2)$, $\beta=2$ for fundamental fermions in 
$SU(N>2)$, and $\beta=4$ in the case of fermions in the adjoint
representation. $N_f$ denotes the number of Dirac or Majorana 
flavors in the cases $\beta=1,2$ and $\beta=4$, respectively. 
$\Sigma_0$ is the magnitude of the quark condensate 
and $f_\pi$ the pion decay constant. The expression (\ref{slope})
summarizes the fact that the spectrum is peaked towards small
virtuality for both $N_f=1$ and $N_{ad}=1$ while it is flat
for $N_f=2$. Effective field theory predicts the slope of the 
Dirac spectrum under the assumption that chiral symmetry is 
broken. The theory cannot predict whether chiral symmetry
breaking takes place for some given $N_f$ or $N_{ad}$.

\section{Conclusions}
\label{sec_con}
 
 In summary we have studied gluino condensation and the 
spectrum of the Dirac operator in an instanton ensemble. 
We employ the semi-classical approximation and focus 
on the Dirac operator in the subspace spanned by the 
zero modes of the individual instantons and anti-instantons.
We have shown how the quaternionic structure of the Dirac
operator in theories with adjoint fermions emerges
naturally from the spin and color structure of the zero
modes. The dependence of the matrix elements on the 
collective coordinates of the instantons is quite 
complicated but qualitatively similar to the simpler 
case of fundamental fermions. 

 We have provided evidence that gluino condensation
does take place in an ensemble of instanton and 
anti-instantons. In a random ensemble, the gluino 
condensate is proportional to the square root of the 
instanton density. In supersymmetric gluodynamics
we find that gluino condensation persists even  if 
interactions between the instantons are taken into 
account. We observed that finite volume effects are 
much weaker than in QCD with two flavors of fundamental 
fermions. This is consistent with the fact that supersymmetric 
gluodynamics has a large mass gap. In QCD with more than one 
adjoint flavor we found no compelling evidence for
gluino condensation.


 There are many problems that remain to be studied. In
particular, it would be interesting to make a systematic
study of gluino and gluino-glueball correlation functions.
There are two types of correlation functions: Constant
correlators that provide information on condensates,
and $x$-dependent correlators related to the spectrum.
These correlation functions will also show to what 
extent supersymmetry is realized in the limit $m_g\to 0$.
In addition to that, it would be interesting to search
for evidence of $Z_2$ domains and to investigate the 
dependence of the results on the topological sector
of the theory. In this work we have used the zero mode 
wave functions that correspond to trivial holonomy and
anti-periodic boundary conditions on the fermions. This 
suggests the question of how the results are changed
if the boundary conditions are modified. In this
case, the zero modes discussed in \cite{CKB_99} will come
into play. Finally, it is important to study the role 
of very large instantons that were excluded in the 
present study.

 There have been suggestions that objects with fractional 
topological charge may play a role in theories with adjoint 
fermions \cite{CG_84,KB_98,KB_98b,DHKM_99}. These objects can
give a direct contribution to the gluino condensate. 
Because of tunneling between the different $Z_N$ phases
the presence of such objects cannot be inferred from the 
behavior of the gluino condensate as a function of the 
quark mass in a finite volume. One should be 
able, however, to detect the presence of fractionally charged
objects in lattice simulations by looking for zero modes of 
the Dirac operator that do not appear in multiples of $2N_c$
\cite{HEN_98}. In this context it would also be interesting 
to study gluino condensation for $N_c>2$. For adjoint fermions 
the number of zero modes per topological charge increases with 
$N_c$. One might therefore doubt that instantons alone are 
sufficient to trigger gluino condensation in large $N_c$
SUSY gluodynamics. It has also been suggested that fractionally 
charged objects can be thought of as instanton constituents 
\cite{FFS_79,KB_98,DM_99}. One might then envision a situation 
where if $N_c$ is small, or instantons are small, fractionally 
charged objects are bound into instantons while for large $N_c$,
or for large instantons, topological objects dissociate and the 
instanton liquid should be replaced by liquid of fractional charges.
 
 Acknowledgments: I would like to thank E. Shuryak, 
J. Verbaarschot and A. Zhitnitsky for useful discussions.
I would also like to thank M. Shifman for many valuable 
comments and pointing out some errors in an earlier version
of this manuscript. This work was supported in part by the 
US DOE grant DE-FG-88ER40388.

\newpage
\appendix
\section{Euclidean Spinor Conventions}
\label{app_conv}

 We use the following euclidean spinor conventions
\bea
&\gamma_\mu &= 
  \left(\begin{array}{rr} 0 & \sigma_\mu \\
                   \bar\sigma_\mu & 0 \end{array}\right)
  = \gamma_\mu^\dagger, \hspace{1cm}
\gamma_5 = \left(\begin{array}{rr} -1 & 0 \\ 
        0 & 1 \end{array}  \right), \\[0.3cm]
&\sigma_\mu &= (i\vec\sigma,1), \hspace{1cm}
  \bar\sigma_\mu = (-i\vec\sigma,1), \\[0.3cm]
&(\sigma_{\mu\nu})_\alpha^{\;\beta} &= \frac{1}{4}
 \left[ (\sigma_\mu)_{\alpha\dot\alpha}
               (\bar\sigma_\nu)^{\dot\alpha\beta}
     -  (\sigma_\nu)_{\alpha\dot\alpha}
               (\bar\sigma_\mu)^{\dot\alpha\beta}\right], \\
&(\bar\sigma_{\mu\nu})^{\dot\alpha}_{\;\dot\beta} &= \frac{1}{4}
 \left[ (\bar\sigma_\mu)^{\dot\alpha\alpha}
               (\sigma_\nu)_{\alpha\dot\beta}
     -  (\bar\sigma_\nu)^{\dot\alpha\alpha}
               (\sigma_\mu)_{\alpha\dot\beta}\right].
\eea
Indices are raised and lowered with $\epsilon^{\alpha\beta}$
and $\epsilon^{\dot\alpha\dot\beta}$ where $\epsilon^{\alpha\beta}
\epsilon_{\beta\gamma}=\delta^\alpha_\gamma$ and $\epsilon^{\dot
\alpha\dot\beta}=\epsilon^{\alpha\beta}$. The euclidean sigma
matrices have the following properties
\bea
(\sigma_\mu\bar\sigma_\nu)_\alpha^{\;\beta} &=&
  \delta_{\mu\nu}\delta_\alpha^\beta 
 + 2(\sigma_{\mu\nu})_\alpha^{\;\beta} ,\\
(\bar\sigma_\mu\sigma_\nu)^{\dot\alpha}_{\;\dot\beta} &=&
  \delta_{\mu\nu}\delta_{\dot\alpha}^{\dot\beta} 
 + 2(\bar\sigma_{\mu\nu})^{\dot\alpha}_{\;\dot\beta} ,\\
(\bar\sigma_\mu)^{\dot\alpha\alpha} &=&
 \epsilon^{\alpha\beta}\epsilon^{\dot\alpha\dot\beta} 
 (\sigma_\mu)_{\beta\dot\beta} \\
\sigma_{\mu\nu} &=& \frac{1}{2}\epsilon_{\mu\nu\rho\sigma}
 \sigma_{\rho\sigma}, 
\hspace{0.5cm}
\bar\sigma_{\mu\nu} = -\frac{1}{2}\epsilon_{\mu\nu\rho\sigma}
 \bar\sigma_{\rho\sigma}.
\eea

\section{Matrix elements}
\label{app_me}

 In this appendix we collect the remaining matrix elements
of the Dirac operator. We define the profile function of the
superconformal zero mode
\be
 \phi_\mu = \frac{\rho}{\sqrt{2}\pi}\frac{x_\mu}
 {(x^2+\rho^2)^2}
\ee
and its Fourier transform $\phi_\mu(k)=-i\hat{k}_\mu\phi_3(k)$
with
\be
\phi_3(k) = -\sqrt{2}\pi\rho^2 K_1(k\rho).
\ee
The matrix elements of the Dirac operator between 
superconformal zero modes are determined by
\be
T^{sc}_\eta(k) = (+i)\left(-2\hat{k}_\eta + 
  8u_\eta(u\cdot\hat{k})^2\right) k |\phi_3(k)|^2 ,
\ee
and the matrix elements between supersymmetric and
superconformal zero modes lead to
\bea
T^{ss-sc}(k) &=& \left(2 - 8 (u\cdot\hat{k})^2\right)
 k\phi_2(k)\phi_3(k),\\
T^{ss-sc}_{\mu\nu}(k) &=& 8\left( u_\mu\hat{k}_\nu
 - u_\nu\hat{k}_\mu \right) (u\cdot\hat{k})
 k\phi_2(k)\phi_3(k).
\eea
From these results we can extract the invariant functions
\be
T^{sc}_1 (z)=-\frac{1}{8\pi^2} \int dk \, 2k^4 j_1(kz)
 |\phi_3(k)|^2
\ee
and $T^{sc}_2(z)=-4T^{sc}_1(z)$ as well as $T^{sc}_3(z)=0$.
Also
\bea
T^{ss-sc}_1(z) &=& \frac{1}{8\pi^2} \int dk \, \left[
 2k^4 j_0(kz) - 8\frac{k^3}{z} j_1(kz) \right]
  \phi_2(k)\phi_3(k) ,\\
T^{ss-sc}_2(z) &=& \frac{1}{8\pi^2} \int dk \,           
 8k^4 j_2(kz) \phi_2(k)\phi_3(k) ,
\eea
and $T^{ss-sc}_3(z)=-T^{ss-sc}_2(z)$. Numerical results for these
functions are shown in Fig. \ref{fig_tia}. The results can be 
parametrized as
\be
\bar\rho T^{sc}_1(z)\;\;\; = \frac{-0.25\bar{z}}{1.0+0.42\bar{z}^2+
 0.21\bar{z}^4}
\ee
as well as
\bea
\bar\rho T^{ss-sc}_1(z) &=&  \frac{-0.17}{1.0+0.05\bar{z}^2
 + 0.08\bar{z}^4}, \\
\bar\rho T^{ss-sc}_2(z) &=&  \frac{1.2\bar{z}^2}{(1.0+0.45\bar{z}^2)^3}
 + \frac{0.014\bar{z}^2}{(1.0+0.21\bar{z}^2)^3}, 
\eea
where $\bar{z}=z/\bar{\rho}$. The overlap matrix elements
$T^{ss-sc}$ are related to the corresponding functions with 
the supersymmetric and superconformal zero modes interchanged.
We find $T^{sc-ss}_{1,2}=-T^{ss-sc}_{1,2}$ and $T^{sc-ss}_3=
T^{ss-sc}_3$.

\newpage\noindent

\begin{figure}
\caption{\label{fig_tia}
Invariant functions characterizing the overlap matrix elements 
of the Dirac operator. Fig. a,b and c show the diagonal overlap 
matrix elements between supersymmetric and superconformal zero 
modes, and the mixed supersymmetric-superconformal matrix
elements.}
\end{figure}

\begin{figure}
\caption{\label{fig_det}
Logarithm of the adjoint fermion determinant in the field of an 
instanton-anti-instanton pair. Fig. a shows $\log\det(\Dslash)$
as a function of $z$ (in units of $\rho$) for $\cos\theta=1$, 
Fig. b gives the dependence on $\cos\theta$ for $z=1$, and
Fig. c the behavior of the determinant as a function of
$\cos\phi$ for $\cos\theta=1/2$ and $z=1$.}
\end{figure}

\begin{figure}
\caption{\label{fig_spec}
Spectrum of the fundamental and adjoint Dirac operator in a
random instanton ensemble. The spectral density is given 
in arbitrary units. }
\end{figure}

\begin{figure}
\caption{\label{fig_qq}
Quark condensate in an interacting instanton ensemble 
as a function of quark or gluino mass. Fig. a shows the 
quark condensate for $N_f=1,\ldots,4$ Dirac fermions in
the fundamental representations $SU(2)$. Fig. b shows the 
gluino condensate for $N_ad=1,2$ Majorana fermions in the 
adjoint representation.}
\end{figure}

\begin{figure}
\caption{\label{fig_speci}
Spectrum of the fundamental and adjoint Dirac operator in an
unquenched instanton ensemble. In the case of the fundamental
spectrum the ensemble was created with $N_f=2$ fundamental 
Dirac fermions, while in the case of the adjoint spectrum
the ensemble corresponds to $N_{ad}=1$ adjoint Majorana 
fermions. }
\end{figure}

\newpage 
\setcounter{figure}{0}

\begin{figure}
\begin{center}
\epsfxsize=10cm
\epsffile{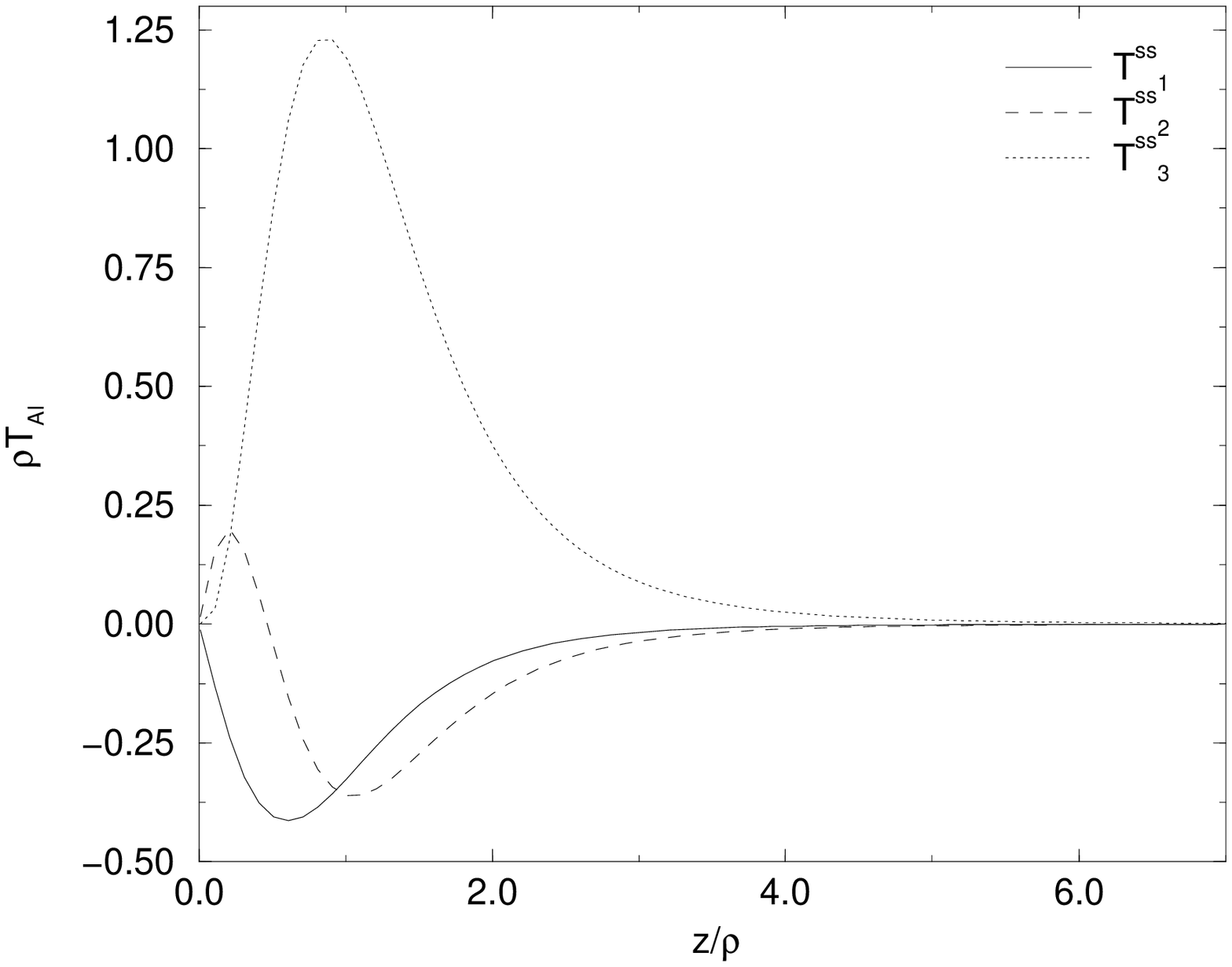}
\end{center}
\begin{center}
\epsfxsize=10cm
\epsffile{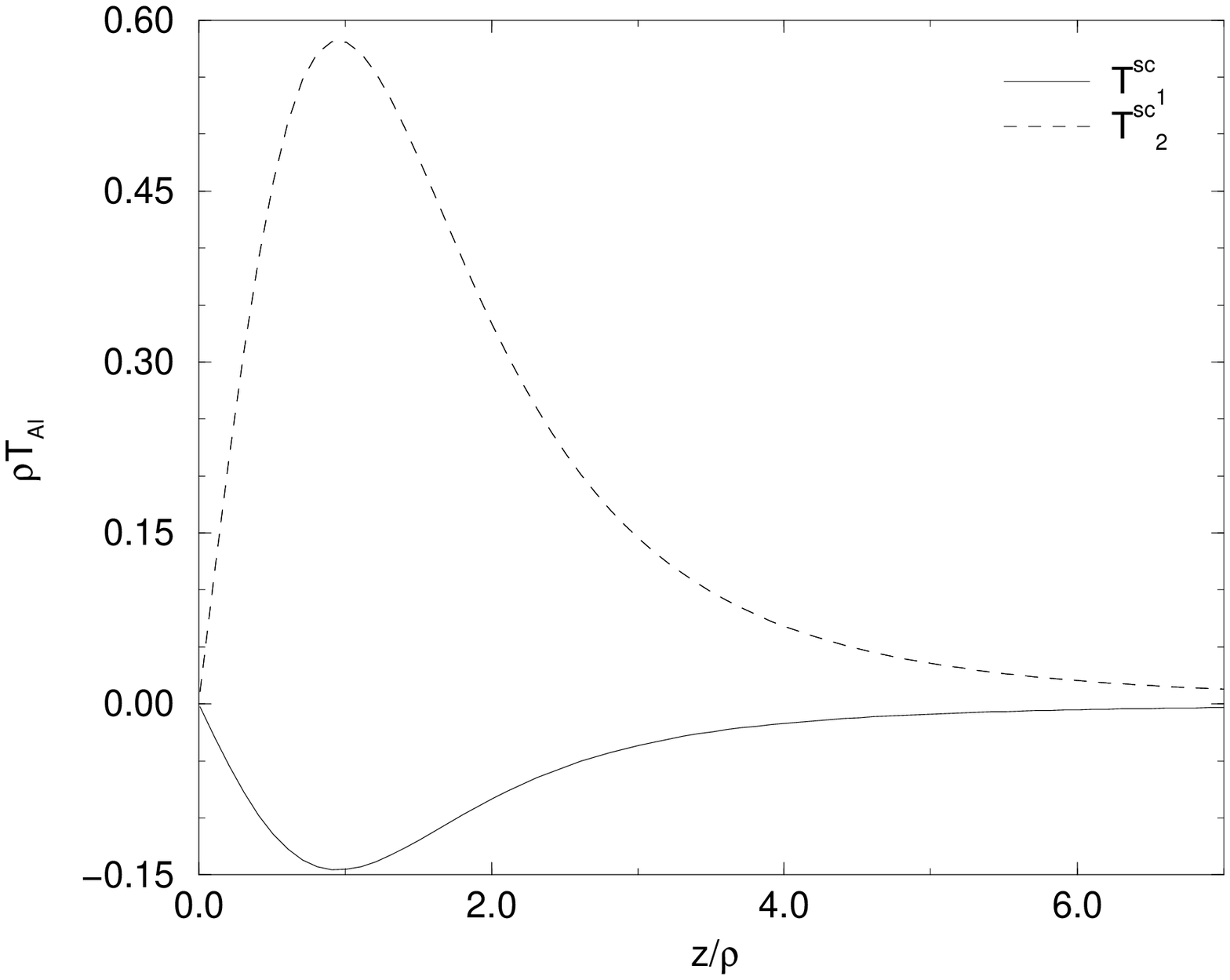}
\end{center}
\begin{center}
\epsfxsize=10cm
\epsffile{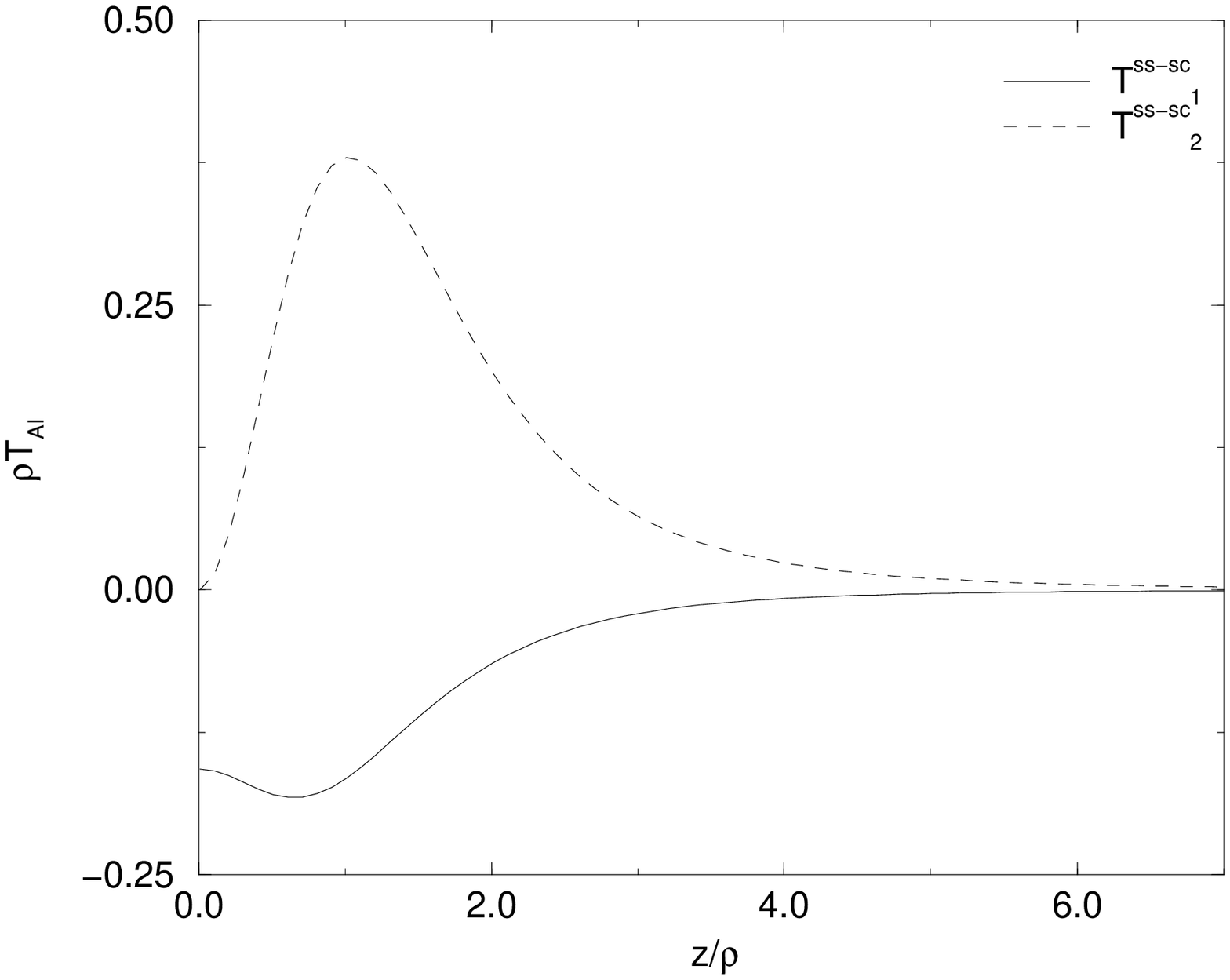}
\end{center}
\caption{}
\end{figure}

\newpage 

\begin{figure}
\begin{center}
\epsfxsize=10cm
\epsffile{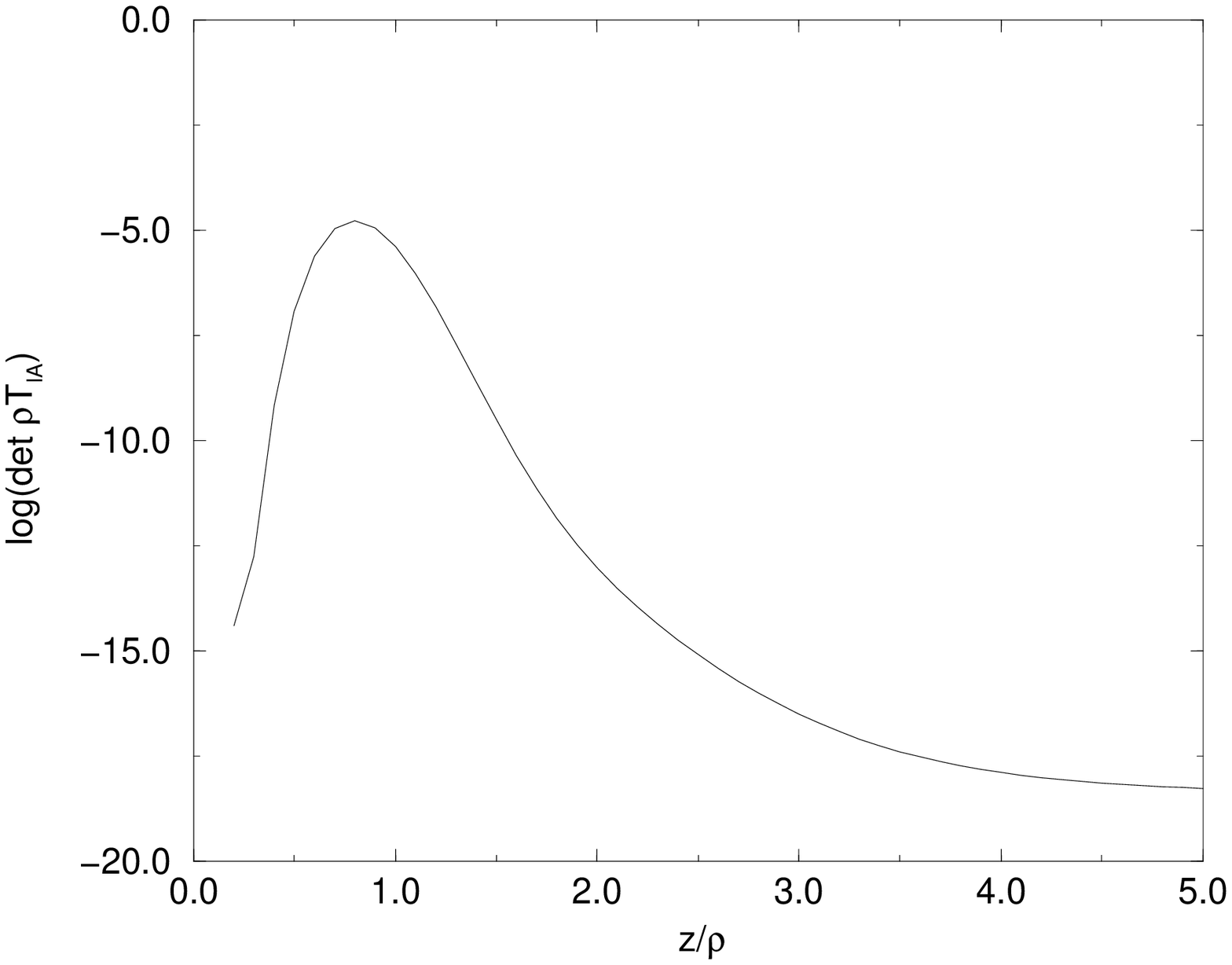}
\end{center}
\begin{center}
\epsfxsize=10cm
\epsffile{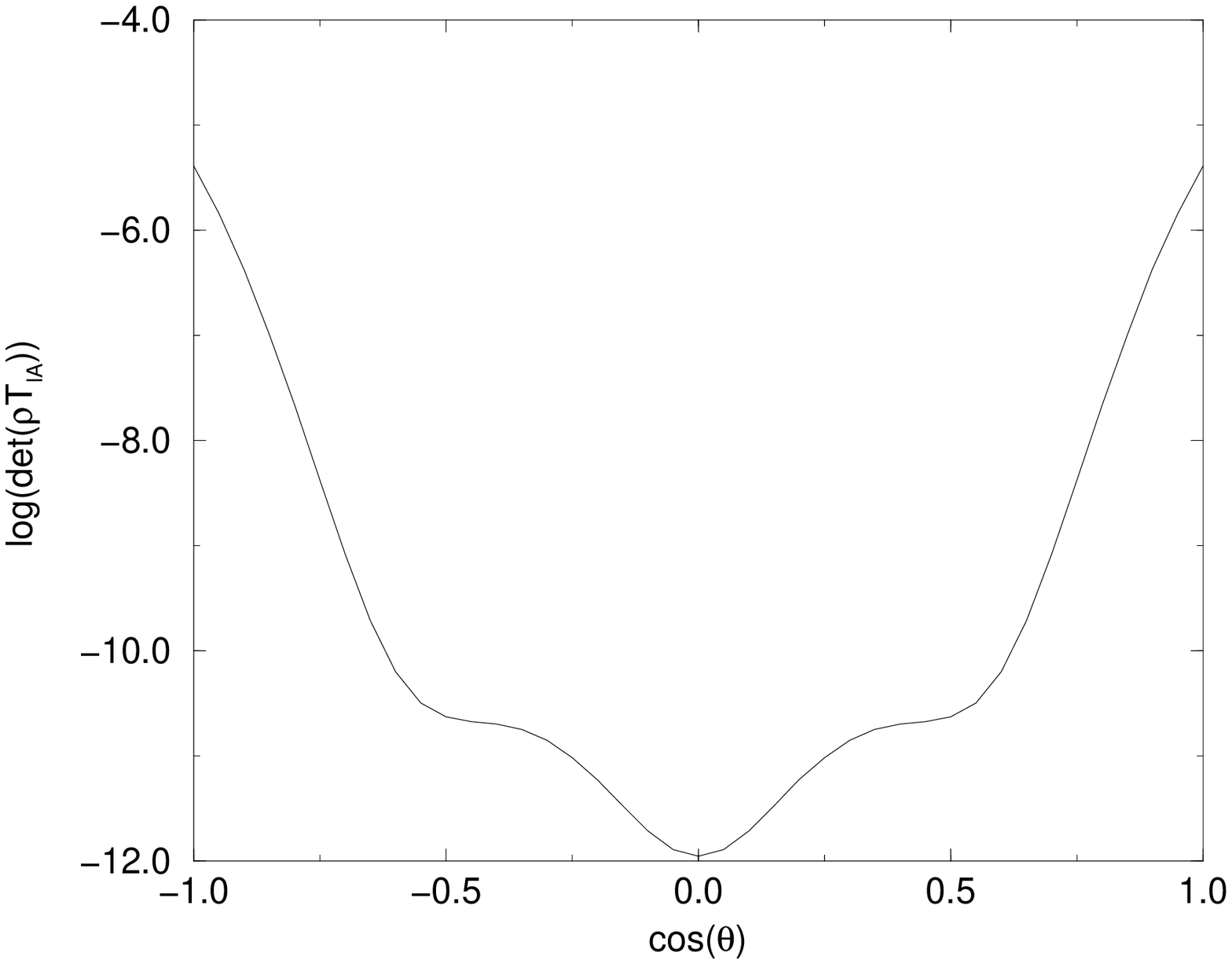}
\end{center}
\begin{center}
\epsfxsize=10cm
\epsffile{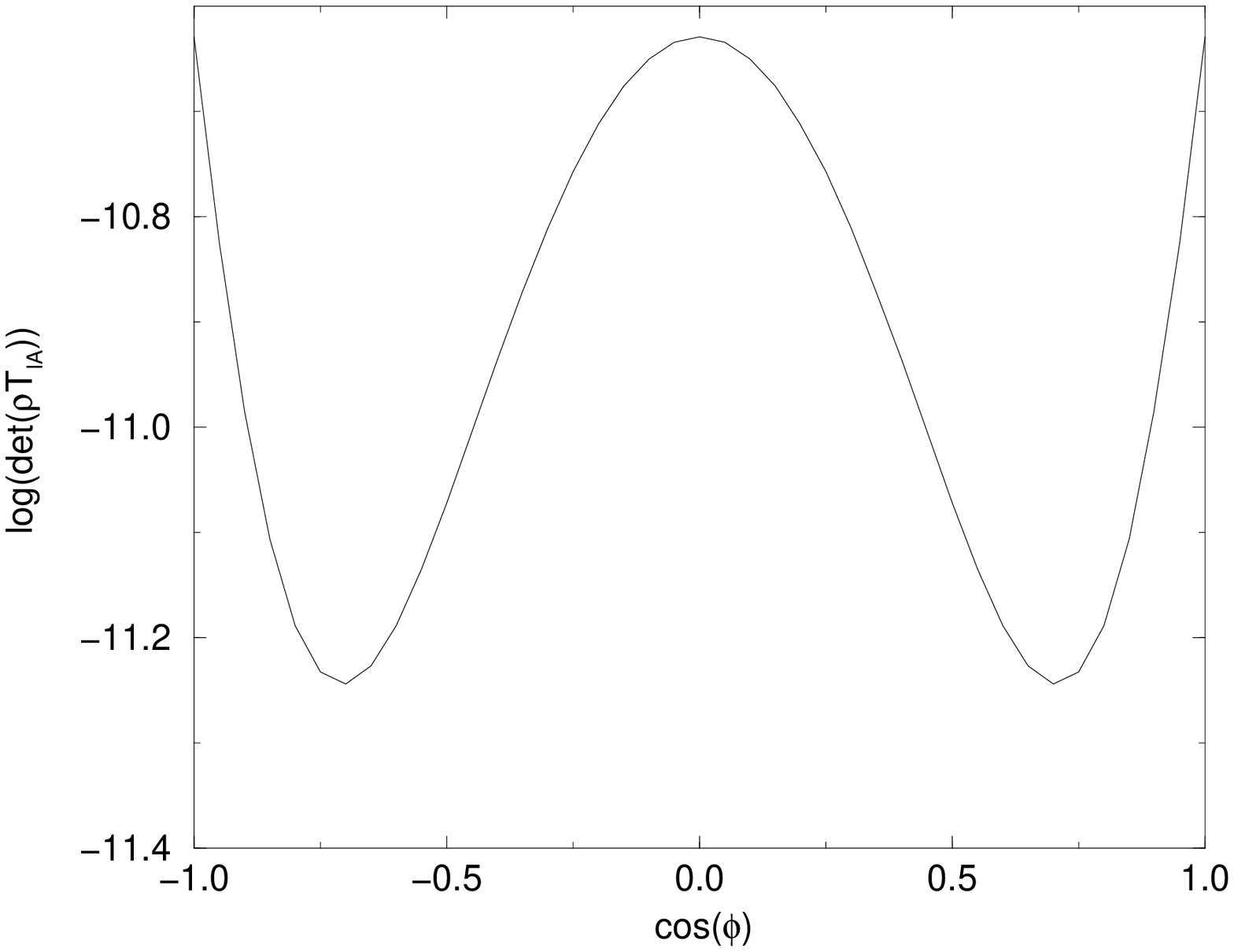}
\end{center}
\caption{}
\end{figure}

\newpage 

\begin{figure}
\begin{center}
\epsfxsize=10cm
\epsffile{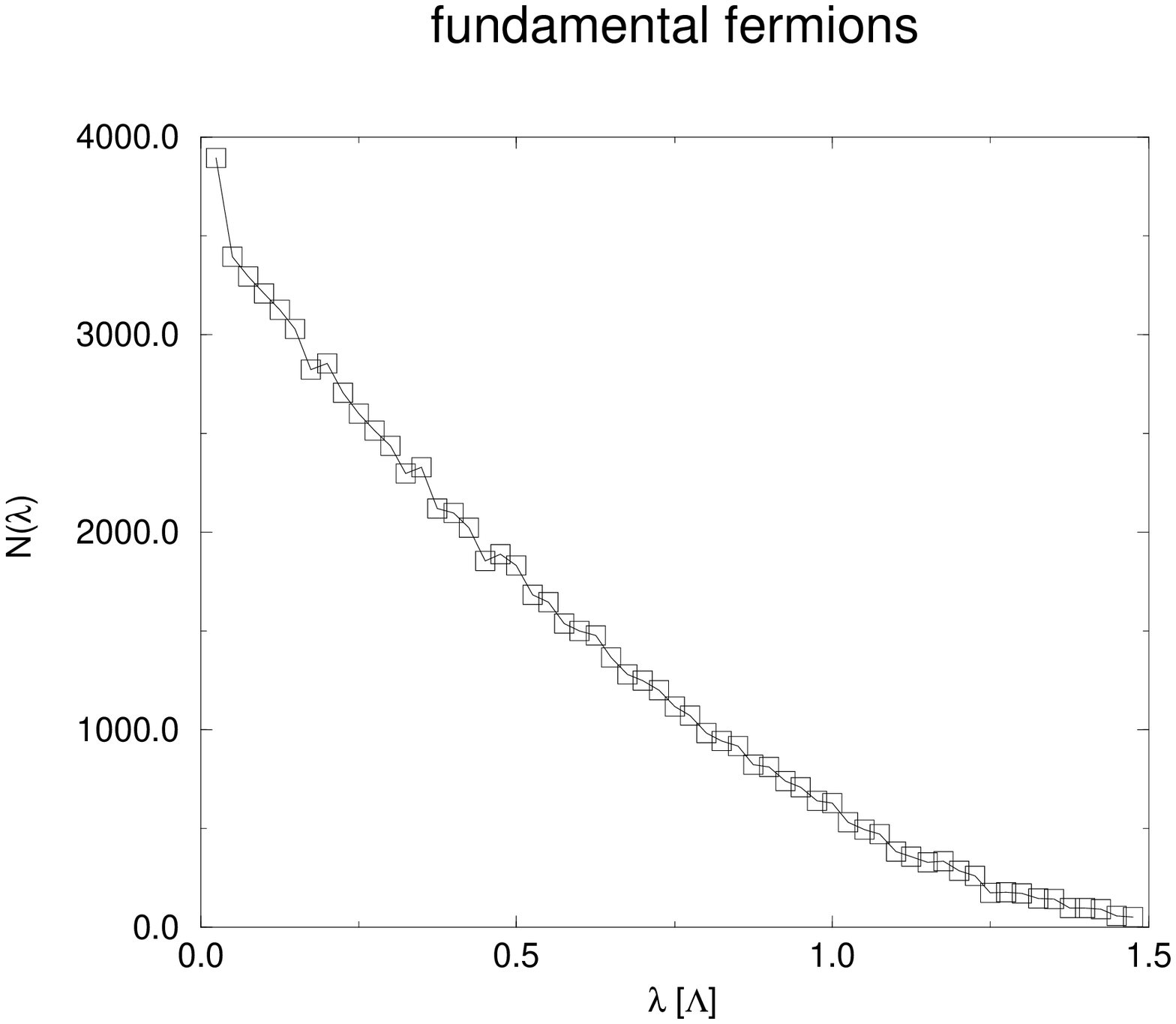}
\end{center}
\begin{center}
\epsfxsize=10cm
\epsffile{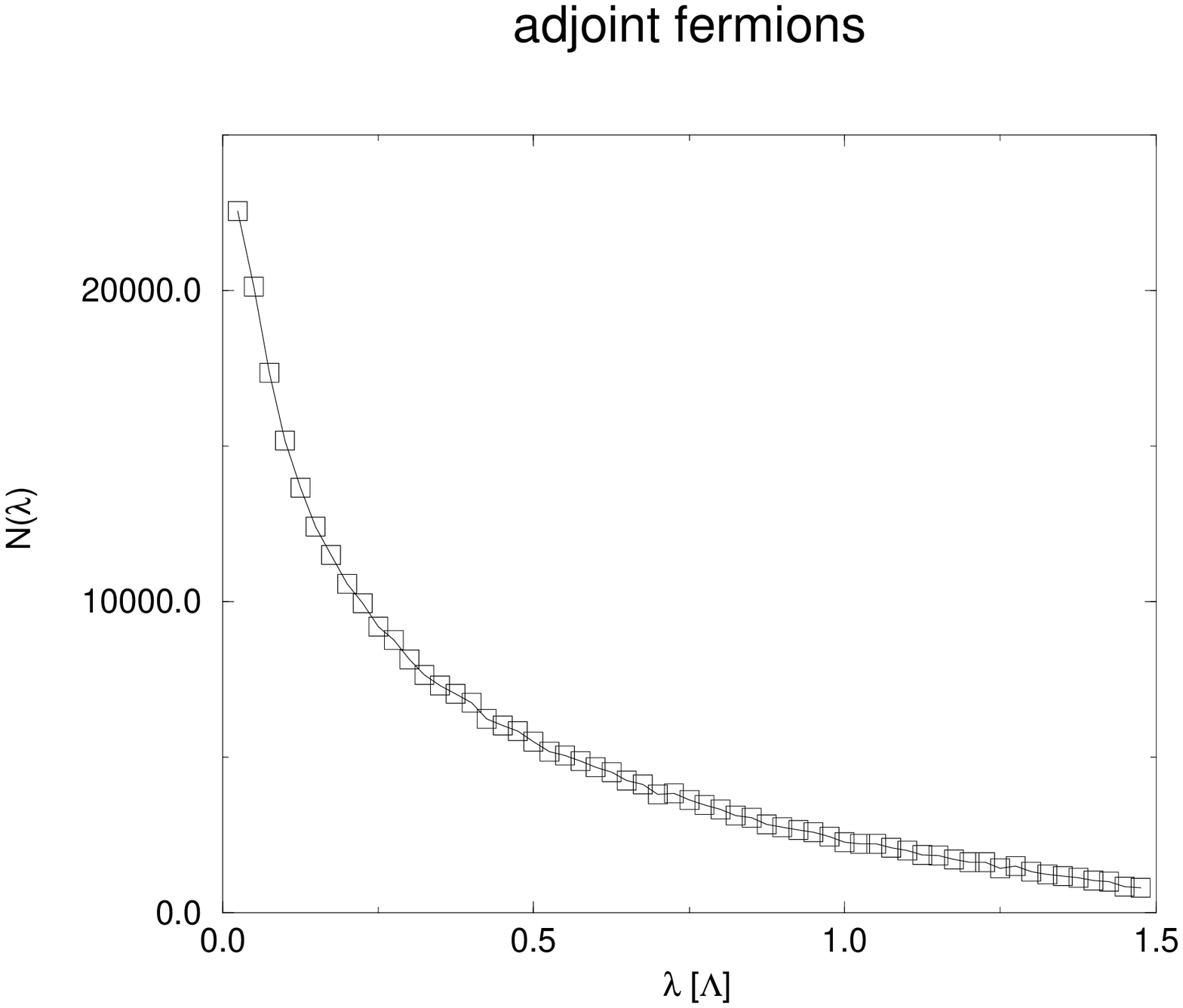}
\end{center}
\caption{}
\end{figure}

\begin{figure}
\begin{center}
\epsfxsize=10cm
\epsffile{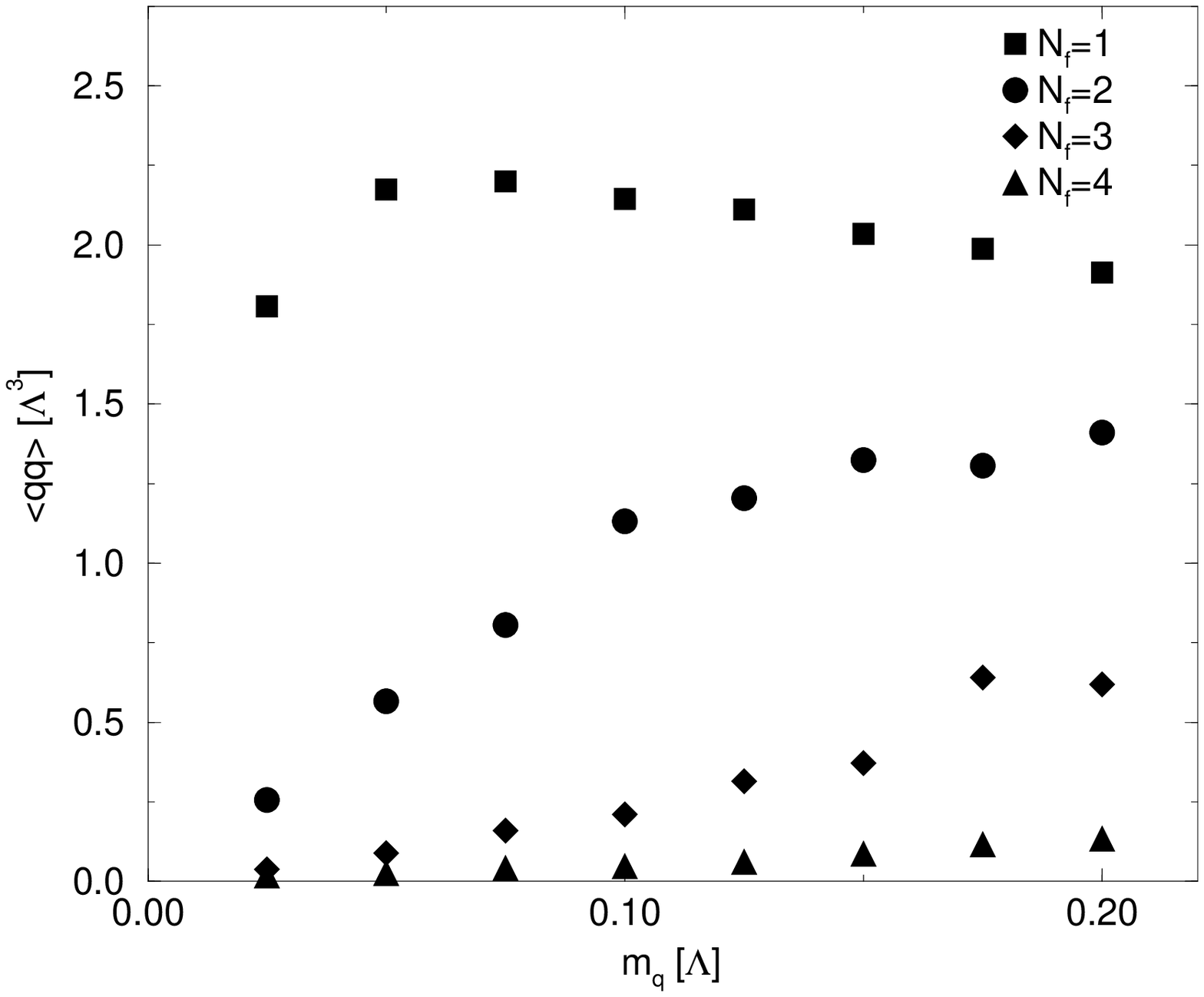}
\end{center}
\begin{center}
\epsfxsize=10cm
\epsffile{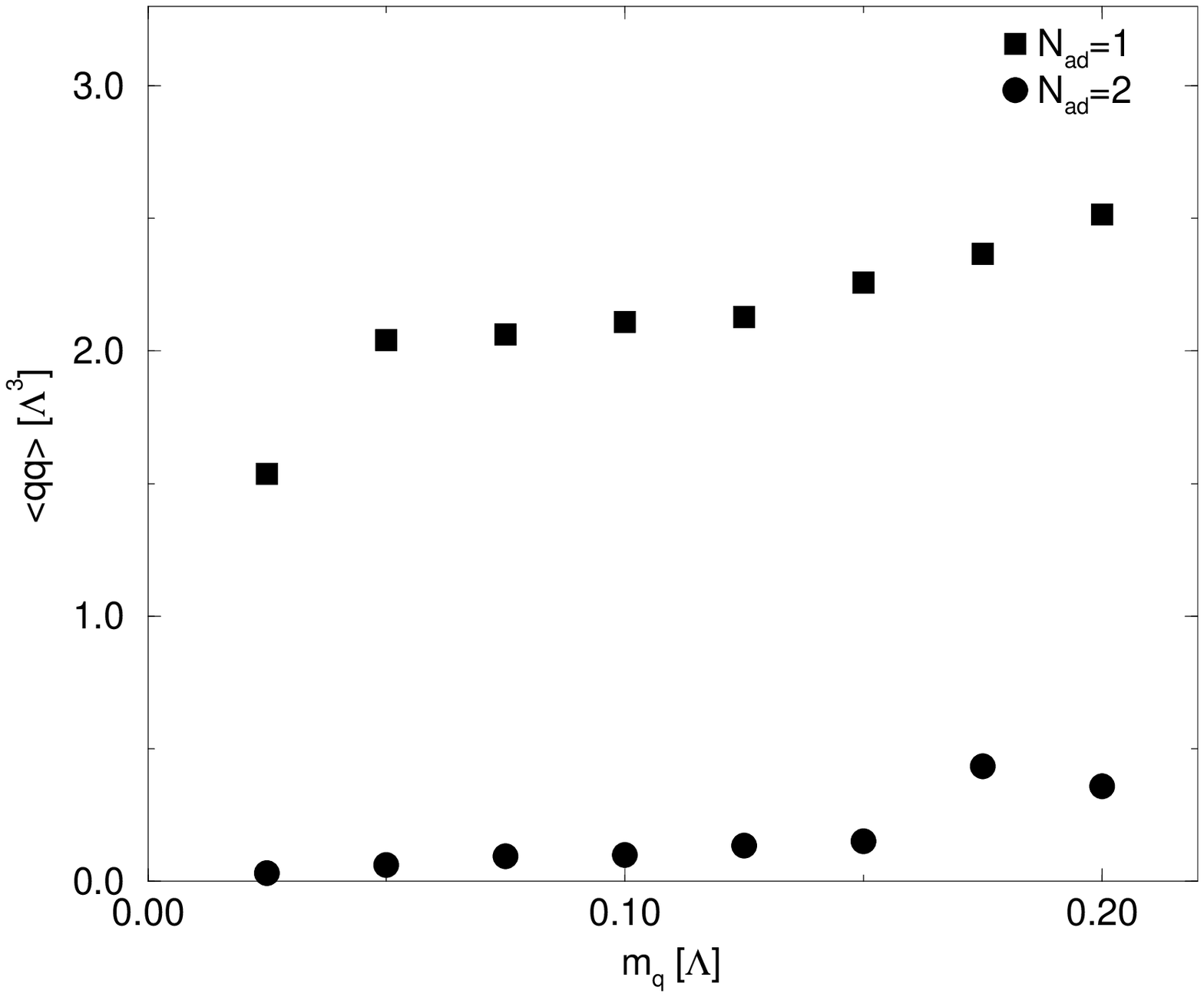}
\end{center}
\caption{}
\end{figure}

\begin{figure}
\begin{center}
\epsfxsize=10cm
\epsffile{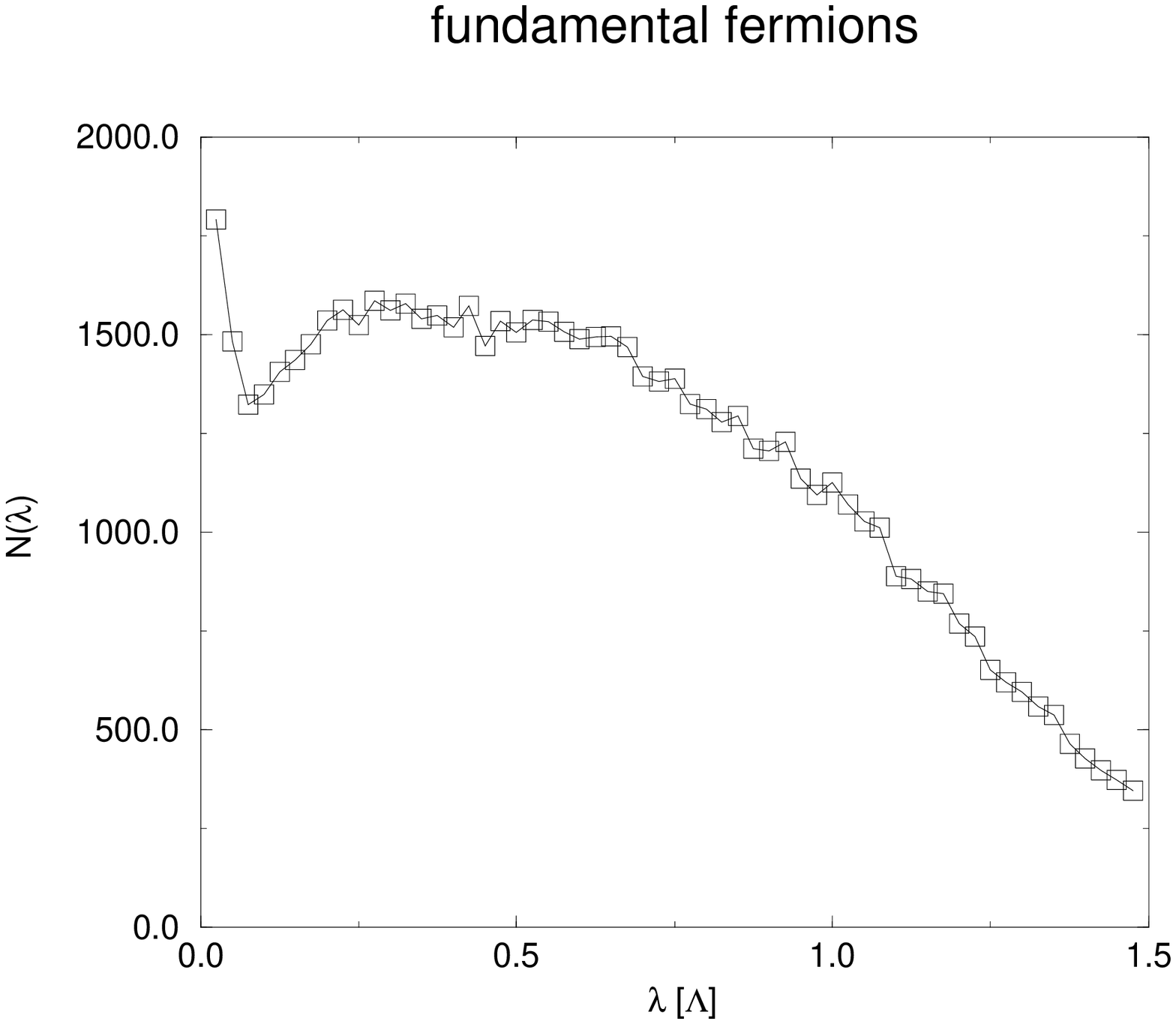}
\end{center}
\begin{center}
\epsfxsize=10cm
\epsffile{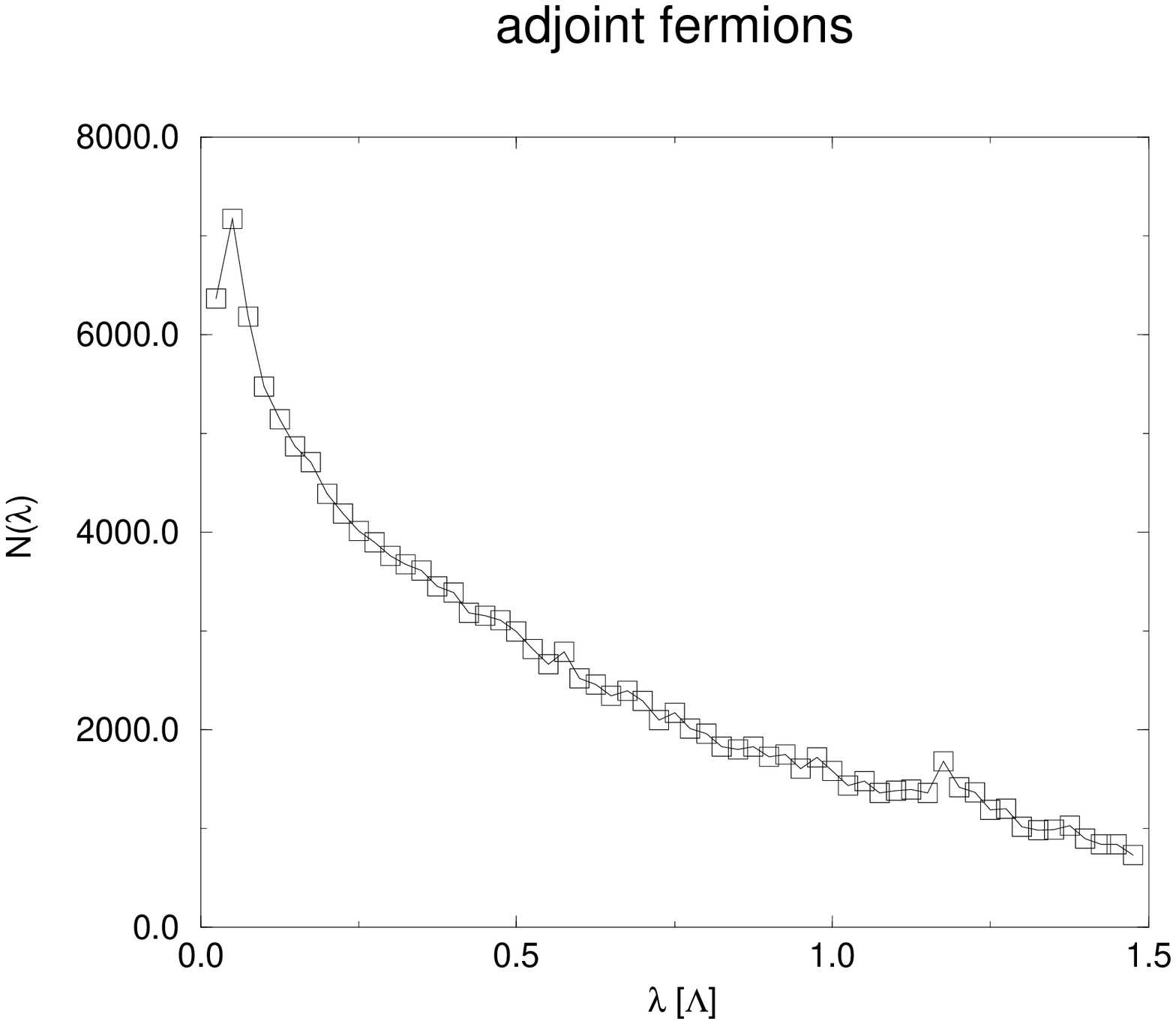}
\end{center}
\caption{}
\end{figure}


\newpage


\begin{thebibliography}{20}

\bibitem{Sei_94}
N.~Seiberg,
Phys.\ Rev.\  {\bf D49} (1994) 6857
[hep-th/9402044].

\bibitem{IS_96}
K.~Intriligator and N.~Seiberg,
Nucl.\ Phys.\ Proc.\ Suppl.\  {\bf 45BC} (1996) 1
[hep-th/9509066].

\bibitem{ADS_83}
I.~Affleck, M.~Dine and N.~Seiberg,
Phys.\ Rev.\ Lett.\  {\bf 51}, 1026 (1983).

\bibitem{ADS_84}
I.~Affleck, M.~Dine and N.~Seiberg,
Nucl.\ Phys.\  {\bf B241}, 493 (1984).

\bibitem{ADS_85}
I.~Affleck, M.~Dine and N.~Seiberg,
Nucl.\ Phys.\  {\bf B256}, 557 (1985).

\bibitem{NSVZ_85}
V.~A.~Novikov, M.~A.~Shifman, A.~I.~Vainshtein and V.~I.~Zakharov,
Nucl.\ Phys.\  {\bf B260}, 157 (1985).

\bibitem{SV_88}
M.~A.~Shifman and A.~I.~Vainshtein,
Nucl.\ Phys.\  {\bf B296}, 445 (1988).

\bibitem{Shi_99}
M.~A.~Shifman and A. Vainshtein,
``Instantons versus Supersymmetry: 15 years later'', 
published in M. Shifman, ITEP lectures in Particle Physics
and Field Theory, Vol. 2, World Scientific, Singapore (1999),
[hep-th/9902018].

\bibitem{NSVZ_83}
V.~A.~Novikov, M.~A.~Shifman, A.~I.~Vainshtein and V.~I.~Zakharov,
Nucl.\ Phys.\  {\bf B229}, 407 (1983).

\bibitem{RV_84}
G.~C.~Rossi and G.~Veneziano,
Phys.\ Lett.\  {\bf B138}, 195 (1984).

\bibitem{ARV_85}
D.~Amati, G.~C.~Rossi and G.~Veneziano,
Nucl.\ Phys.\  {\bf B249}, 1 (1985).

\bibitem{FS_86}
J.~Fuchs and M.~G.~Schmidt,
Z.\ Phys.\  {\bf C30}, 161 (1986).

\bibitem{AKM*88}
D.~Amati, K.~Konishi, Y.~Meurice, G.~C.~Rossi and G.~Veneziano,
Phys.\ Rept.\  {\bf 162}, 169 (1988).

\bibitem{KS_97}
A.~Kovner and M.~Shifman,
Phys.\ Rev.\  {\bf D56}, 2396 (1997)
[hep-th/9702174].

\bibitem{DHKM_99}
N.~M.~Davies, T.~J.~Hollowood, V.~V.~Khoze and M.~P.~Mattis,
Nucl.\ Phys.\  {\bf B559}, 123 (1999)
[hep-th/9905015].

\bibitem{HKLM_99}
T.~J.~Hollowood, V.~V.~Khoze, W.~Lee and M.~P.~Mattis,
hep-th/9904116.

\bibitem{VR_99}
A.~Ritz and A.~Vainshtein, 
Nucl.\ Phys.\ {\bf B566}, 311 (2000)
[hep-th/9909073].

\bibitem{Shu_82}
E.~V.~Shuryak,
Nucl.\ Phys.\ {\bf B203}, 93 (1982).

\bibitem{DP_85}
D.~I.~Diakonov and V.~Y.~Petrov,
Sov.\ Phys.\ JETP {\bf 62}, 204 (1985).

\bibitem{DP_86}
D.~I.~Diakonov and V.~Y.~Petrov,
Nucl.\ Phys.\ {\bf B272}, 457 (1986).

\bibitem{BC_80}
T.~Banks and A.~Casher,
Nucl.\ Phys.\ {\bf B169}, 103 (1980).

\bibitem{SS_98} 
T. Sch{\"a}fer and E. V. Shuryak, 
Rev. Mod. Phys. {\bf 70}, 323 (1998).

\bibitem{Neg_99}
J.~W.~Negele,
Nucl.\ Phys.\ Proc.\ Suppl.\  {\bf 73}, 92 (1999)
[hep-lat/9810053].

\bibitem{KSV_85}
Y.~I.~Kogan, M.~A.~Shifman and M.~I.~Vysotsky,
Sov.\ J.\ Nucl.\ Phys.\  {\bf 42}, 318 (1985).

\bibitem{LS_92}
H.~Leutwyler and A.~Smilga,
Phys.\ Rev.\  {\bf D46}, 5607 (1992).

\bibitem{VY_82}
G.~Veneziano and S.~Yankielowicz,
Phys.\ Lett.\  {\bf B113}, 231 (1982).

\bibitem{tHo_76}
G.~'t Hooft,
Phys.\ Rev.\  {\bf D14}, 3432 (1976).

\bibitem{VZ_82}
A.~I.~Vainshtein and V.~I.~Zakharov,
JETP Lett.\  {\bf 35}, 323 (1982).

\bibitem{NSVVZ_83}
V.~A.~Novikov, M.~A.~Shifman, A.~I.~Vainshtein, M.~B.~Voloshin 
and V.~I.~Zakharov,
Nucl.\ Phys.\  {\bf B229}, 394 (1983).

\bibitem{Ver_94}
J.~J.~M.~Verbaarschot,
Phys.\ Rev.\ Lett.\  {\bf 72}, 2531 (1994)
[hep-th/9401059].

\bibitem{Ver_92}
J.~J.~M.~Verbaarschot,
Nucl.\ Phys.\  {\bf B362}, 33 (1992).

\bibitem{CDG_78}
C.~G.~Callan, R.~Dashen and D.~J.~Gross,
Phys.\ Rev.\  {\bf D17}, 2717 (1978).

\bibitem{OV_98}
J.~C.~Osborn and J.~J.~Verbaarschot,
Nucl.\ Phys.\  {\bf B525}, 738 (1998)
[hep-ph/9803419].

\bibitem{NSVZ_83b}
V.~A.~Novikov, M.~A.~Shifman, A.~I.~Vainshtein and V.~I.~Zakharov,
Nucl.\ Phys.\  {\bf B229}, 381 (1983).

\bibitem{Cam_99}
I.~Campos {\it et al.}  [DESY-Munster Collaboration],
Eur.\ Phys.\ J.\  {\bf C11}, 507 (1999)
[hep-lat/9903014].

\bibitem{Kir_99}
R.~Kirchner, I.~Montvay, J.~Westphalen, S.~Luckmann and K.~Spanderen  
[DESY-Munster Collaboration],
Phys.\ Lett.\  {\bf B446}, 209 (1999)
[hep-lat/9810062].

\bibitem{SS_93}
A.~V.~Smilga and J.~Stern,
Phys.\ Lett.\  {\bf B318}, 531 (1993).

\bibitem{TV_99}
D.~Toublan and J.~J.~Verbaarschot,
Nucl.\ Phys.\  {\bf B560}, 259 (1999)
[hep-th/9904199].

\bibitem{CKB_99}
M.~N.~Chernodub, T.~C.~Kraan and P.~van Baal,
hep-lat/9907001.

\bibitem{CG_84}
E.~Cohen and C.~Gomez,
Phys.\ Rev.\ Lett.\  {\bf 52}, 237 (1984).

\bibitem{KB_98}
T.~C.~Kraan and P.~van Baal,
Phys.\ Lett.\  {\bf B435}, 389 (1998)
[hep-th/9806034].

\bibitem{KB_98b}
T.~C.~Kraan and P.~van Baal,
Nucl.\ Phys.\  {\bf B533}, 627 (1998)
[hep-th/9805168].

\bibitem{HEN_98}
U.~M.~Heller, R.~G.~Edwards and R.~Narayanan,
Nucl.\ Phys.\ Proc.\ Suppl.\  {\bf 73}, 497 (1999)
[hep-lat/9810003].

\bibitem{FFS_79}
V.~A.~Fateev, I.~V.~Frolov and A.~S.~Shvarts,
Nucl.\ Phys.\  {\bf B154}, 1 (1979).

\bibitem{DM_99}
D.~Diakonov and M.~Maul,
hep-th/9909078.


\end{thebibliography}
\end{document}